\documentclass[twocolumn]{aastex631}
\usepackage{amsmath, textgreek, enumitem, booktabs, needspace, float, xcolor}

\received{April 1, 2025}
\revised{May 6, 2025}
\accepted{May 18, 2025}

\submitjournal{ApJL}

\shorttitle{The Receding Cosmic Shoreline of Mid-to-Late M Dwarfs}
\shortauthors{Pass, Charbonneau, \& Vanderburg}

\begin{document}
\title{The Receding Cosmic Shoreline of Mid-to-Late M Dwarfs: \\ Measurements of Active Lifetimes Worsen Challenges for Atmosphere Retention by Rocky Exoplanets}

\author[0000-0002-1533-9029]{Emily K. Pass}
\altaffiliation{Juan Carlos Torres Postdoctoral Fellow}
\affiliation{Kavli Institute for Astrophysics and Space Research, Massachusetts Institute of Technology, Cambridge, MA 02139, USA}

\author[0000-0002-9003-484X]{David Charbonneau}
\affiliation{Center for Astrophysics $\vert$ Harvard \& Smithsonian, 60 Garden Street, Cambridge, MA 02138, USA}

\author[0000-0001-7246-5438]{Andrew Vanderburg}
\altaffiliation{Sloan Research Fellow}
\affiliation{Kavli Institute for Astrophysics and Space Research, Massachusetts Institute of Technology, Cambridge, MA 02139, USA}

\begin{abstract}
\noindent Detecting and characterizing the atmospheres of terrestrial exoplanets is a key goal of exoplanetary astronomy, one that may now be within reach given the upcoming campaign to conduct a large-scale survey of rocky M-dwarf worlds with the James Webb Space Telescope. It is imperative that we understand where known planets sit relative to the cosmic shoreline, the boundary between planets that have retained atmospheres and those that have not. Previous works modeled the historic XUV radiation received by mid-to-late M-dwarf planets using a scaling relation calibrated using more massive stars, but fully convective M dwarfs display unique rotation/activity histories that differ from Sun-like stars and early M dwarfs. We synthesize observations of the active lifetimes of mid-to-late M dwarfs to present an updated estimate of their historic XUV fluence. For known planets of inactive, mid-to-late M dwarfs, we calculate a historic XUV fluence that is 2.1--3.1 times the canonical XUV scaling relation on average, with the larger value including corrections for the pre-main-sequence phase and energetic flares. We find that only the largest terrestrial planets known to orbit mid-to-late M-dwarfs are likely to have retained atmospheres within the cosmic shoreline paradigm. Our calculations may help to guide the selection of targets for JWST and may prove useful in interpreting the results; to this end, we define a novel Atmosphere Retention Metric (ARM) that indicates the distance between a planet and the cosmic shoreline, and tabulate the ARM for known mid-to-late M-dwarf planets.
\end{abstract}

\section{Introduction} \label{sec:intro}

Planetary atmospheres are sculpted by stellar activity. \citet{Zahnle2017} introduce the concept of the ``cosmic shoreline,'' in which the population of planets with and without atmospheres can be understood as the end result of atmospheric escape. The relevant parameters in this framework are instellation and planetary escape velocity. However, knowledge of the present-day instellation is not enough: one requires the cumulative historic XUV irradiation received by the planet, $I_{\rm XUV}$. This parameter cannot be measured directly. A commonly used prescription for an M dwarf's XUV history is a saturated regime followed by a gradual power-law decrease with age \citep[e.g.,][]{Selsis2007, Luger2015, Ribas2016, Moore2020, Engle2024}, motivated by the rotation--activity relation and the rotational evolution observed for Sun-like stars. Here, we argue that such a treatment is insufficient to accurately predict the location of mid-to-late M-dwarf rocky planets relative to the cosmic shoreline.

\begin{figure*}
    \includegraphics[width=\textwidth]{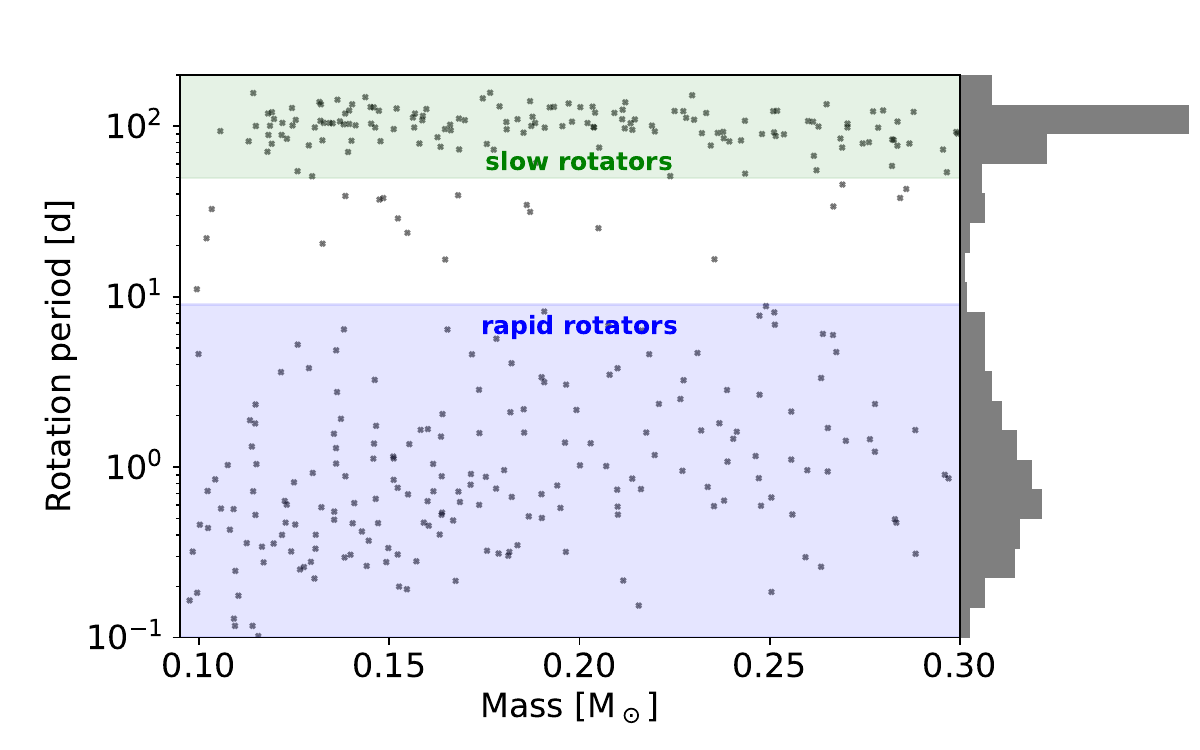}
    \caption{Rotation periods of $<$0.3M$_\odot$ M dwarfs in the field, as observed by the MEarth array \citep{Nutzman2008, Irwin2015} in \citet{Newton2016, Newton2018}, with masses revised using Gaia DR3 parallaxes \citep{Gaia2023} and the \citet{Benedict2016} $K$-band mass--luminosity relation. Probable unresolved binaries have been removed based on Gaia astrometric indicators. The field population is bimodal: there are rapid rotators with rotation periods of roughly a day and slow rotators with periods of roughly 100 days. Very few stars fall between the modes.}
    \label{fig:rot}
\end{figure*}

To briefly summarize the activity--rotation relation, stars are born with angular momentum inherited from their birth cloud, and they spin up further as they accrete material and contract onto the main sequence. Over time, that angular momentum is lost through magnetized stellar winds, causing the star to spin down. Magnetic activity and rotation are thus closely linked: this relationship is typically quantified in terms of the Rossby number, $\rm{Ro}\equiv\mathit{P}_{\rm rot}/\tau_c$, where $P_{\rm rot}$ is the stellar rotation period and $\tau_{\rm{c}}$ is the convective turnover time \citep{Noyes1984}. This relation exhibits the same broken-power-law behavior in various activity indicators, including X-ray emission \citep{Wright2011, Wright2018, Engle2024}, H\textalpha\ emission \citep{Newton2017}, UV emission \hbox{\citep{France2018},} and flare rate \citep{Medina2020, Medina2022_flare}. At short periods, stars have a high activity level that does not depend on rotation rate; this is the saturated regime.  As the stellar rotation period increases, the stars enter the unsaturated regime and the activity decreases. The origin of this saturation remains unclear \citep{Wright2011, Wright2018}; possible explanations include saturation of the dynamo, saturation of the active-region filling factor, or centrifugal stripping of the corona \citep{Vilhu1984, Jardine1999}.

While the rotation--activity relation has been shown to be surprisingly universal across spectral types \citep{Newton2017, Wright2018, Lehtinen2020}, the same is not true for rotational evolution. While Sun-like stars spin down gradually, following the Skumanich law \citep[$\omega \propto t^{-1/2}$;][]{Skumanich1972}, mid-to-late M dwarfs do not. Rather, these stars make an abrupt transition from rapid to slow rotation. This inference is based on the bimodal distribution of rotation periods for low-mass M dwarfs in the field (Figure~\ref{fig:rot}). Many of these stars rotate rapidly, many others rotate slowly, and few rotate with intermediate periods. As stars in our solar neighborhood have a roughly uniform distribution of ages \citep[e.g.,][]{Fantin2019}, the gap between the modes implies that low-mass M dwarfs cannot be experiencing Skumanich-law spindown; otherwise, the gap would be completely filled in with stars.\footnote{Note that the dearth of stars in the gap \textbf{cannot} simply be explained by rotation periods being more difficult to detect for intermediate rotators (for example, due to a difference in the coherence of spot patterns): as illustrated in \citet{Pass2023_active}, the bimodality of Figure~\ref{fig:rot} holds even if we construct a volume-complete sample and successfully detect rotation periods for all stars exhibiting H\textalpha\ emission, which is a reliable proxy for rotation \citep{Newton2017}.} At some point in these stars' lives, magnetic braking must increase dramatically, causing them to rapidly shed angular momentum and transition to a slowly rotating state; they therefore spend very little of their lives rotating with periods of $10<P_{\rm rot}<50$ days. Correspondingly, they spend little time with intermediate activity levels, transitioning quickly between the high activity levels of the saturated regime and the low activity levels associated with long periods (a decline of roughly two orders of magnitude). A proper accounting of this evolutionary history is necessary to accurately predict which mid-to-late M-dwarf planets are able to retain atmospheres. 

In contrast, the XUV prescription adopted by \citet{Zahnle2017} has been often employed in the literature, and so we reproduce the relevant equations here. Drawing from the work of \citet{Penz2008a} and \citet{Penz2008b}, which fit scaling relations to ROSAT \citep{Truemper1982} observations of the Pleiades and Hyades clusters, \citet{Lammer2009} present equations for the time-dependent X-ray luminosity of F-, G-, K-, and M-dwarf stars. For M-dwarfs, this equation takes the form
\begin{equation}\label{eq:LX_M}
    L_{\rm X}(t) = \begin{cases} 
          0.17 L_0 t^{-0.77} & t\leq 0.6 \rm{Gyr} \\
          0.13 L_0 t^{-1.34} & t > 0.6 \rm{Gyr}
       \end{cases}, L_0 = 10^{28.75},
\end{equation}

\noindent while for G-dwarfs like the Sun,
\begin{equation} \label{eq:LX_G}
    L_{\rm X}(t) = \begin{cases} 
          0.375 L_0 t^{-0.425} & t\leq 0.6 \rm{Gyr} \\
          0.19 L_0 t^{-1.69} & t > 0.6 \rm{Gyr}
       \end{cases}, L_0 = 10^{29.35},
\end{equation}

\noindent with $t$ in Gyr and $L$ in erg s$^{-1}$. As the ROSAT measurements cover the transition region between the soft X-ray and the EUV and because they correlate well with EUVE
satellite measurements from \citet{Ribas2005}, \citet{Lammer2009} further presume that these scaling relations can be treated as a ``justified EUV proxy''; i.e., these scalings are representative for the full XUV region of interest. \citet{Zahnle2017} then integrate these equations and approximate the results across spectral types as a simple power law,
\begin{equation}
L_{\rm XUV} \propto L_{\rm bol}^{0.4},
\end{equation}

\noindent or in terms of the historic XUV instellation of a planet with semimajor axis $a$,
\begin{equation} \label{eq:Ixuv}
\frac{I_{\rm XUV}}{I_{\rm XUV, \oplus}} = \frac{a_\oplus^2}{a^2} \Big( \frac{L_{\rm bol}}{L_\odot} \Big)^{0.4}.
\end{equation}

The \citet{Zahnle2017} cosmic shoreline paradigm is a simplification of the complex processes that govern atmospheric escape, and various assumptions are open to debate. One point of contention is whether the ROSAT observations are truly a justified EUV proxy \citep[e.g.,][]{King2021, Johnstone2021}. This work focuses on another concern: the fact that Equation~\ref{eq:LX_M} is inconsistent with our current understanding of the activity evolution of mid-to-late M dwarfs, as outlined above. The true XUV history of these stars should include an abrupt transition between rapid rotation / magnetic activity and slow rotation / magnetic quiescence. Furthermore, mid-to-late M dwarfs remain in the saturated regime for much longer than 0.6Gyr, although this duration is highly mass dependent: \citet{Pass2024} find an average active lifetime of 1.3Gyr for M dwarfs with masses of 0.3M$_\odot$, 2.8Gyr for 0.2M$_\odot$, and 4.4Gyr for 0.1M$_\odot$. It is therefore inadvisable to use Equation~\ref{eq:Ixuv} to infer the location of exoplanets relative to the cosmic shoreline for mid-to-late M dwarfs, although it has nonetheless been used for that purpose in the literature.

How inaccurate is this approximation for the planets of small stars? This work seeks to answer that question, applying our current knowledge of the activity evolution of mid-to-late M dwarfs to the population of known planets orbiting these stars and providing an updated prediction of the cosmic shoreline for mid-to-late M dwarfs. Such a revision is timely given ongoing efforts to use the James Webb Space Telescope to map the cosmic shoreline for M dwarfs, and in particular, a recent program consisting of 500 hours of Director's Discretionary time to search for atmospheres of rocky M-dwarf planets \citep{Redfield2024}. Notably, the Targets Under Consideration (TUC) Version 1.0 list for this program currently uses the \cite{Zahnle2017} scaling relation to prioritize targets. By more appropriately modeling the XUV history of these small stars, we can better predict which terrestrial planets may be capable of retaining an atmosphere, thus identifying the targets that should be highest priority for accomplishing the scientific mission of this JWST survey. Our predictions also provide a testable hypothesis that can assist in the interpretation of the results of such a program.

\section{Methods}

An estimate of the $I_{\rm XUV}$ fluence received by mid-to-late M-dwarf terrestrial planets requires two key parameters: $\big(\frac{L_{\rm X}}{L_{\rm bol}}\big)_{\rm sat}$, the fractional X-ray luminosity of the star during the saturated phase, and $t_{\rm sat}$, the duration of that phase. In this section, we first identify appropriate choices for these parameters. In Section~\ref{sec:results}, we then apply the resultant relation to the sample of mid-to-late M dwarfs with known terrestrial planets.

\subsection{The X-ray fluence during saturation}

For $\big(\frac{L_{\rm X}}{L_{\rm bol}}\big)_{\rm sat}$, we can turn to \citet{Wright2018}, who investigated the rotation--activity relationship for fully convective M dwarfs using X-ray luminosities in the ROSAT band. They modeled the saturated/unsaturated regimes using the prescription
\begin{equation}
\label{eq:saturate}
    \frac{L_{\rm X}}{L_{\rm bol}} = \begin{cases} 
          C \rm {Ro}^\beta & \rm{Ro} > \rm{Ro}_{\rm sat}\\
          \big(\frac{L_{\rm X}}{L_{\rm bol}}\big)_{\rm sat} & \rm{Ro} \leq \rm{Ro}_{\rm sat}
       \end{cases},
\end{equation}

\noindent finding best-fit values of $\beta$=$-2.3^{+0.4}_{-0.6}$, $\rm{Ro}_{\rm sat} = 0.14^{+0.08}_{-0.04}$, and log$\big(\frac{L_{\rm X}}{L_{\rm bol}}\big)_{\rm sat}=-3.05^{+0.05}_{-0.06}$.

Note that a Rossby number of Ro=0.14 corresponds to a rotation period of 10--20 days for 0.1--0.3M$_\odot$ M dwarfs. Other work finds that the transition between regimes occurs at a somewhat larger Rossby number, such as Ro=0.2 \citep[from H\textalpha;][]{Newton2017} or Ro=0.5 \citep[from flares;][]{Medina2022_flare}, the latter corresponding to rotation periods of 30--80 days. This inconsistency may stem from the rarity of these intermediate rotators, which causes the Rossby number of the transition to be easily biased by outliers, or differences in when the transition occurs for coronal vs chromospheric phenomena \citep{Linsky2020}, but its precise value is immaterial for our purposes: regardless, the transition from saturation to unsaturation occurs during the gap between rotation modes. A 0.2M$_\odot$ M dwarf rotating with a period of 100 days (the locus of the slowly rotating mode) has Ro=1.1, representing a decrease in $\frac{L_{\rm X}}{L_{\rm bol}}$ of more than two orders of magnitude. In summary: until $t_{\rm sat}$, a mid-to-late M dwarf emits $\frac{L_{\rm X}}{L_{\rm bol}}$ = 10$^{-3.05}$. After $t_{\rm sat}$, the XUV fluence is negligible. The equation for $I_{\rm XUV}$ is therefore given by
\begin{equation}\label{eq:our_Ixuv}
I_{\rm XUV} = \int_{0}^{t_{\rm sat}} \frac{L_{\rm X}}{L_{\rm bol}} \frac{L_{\rm bol}}{4\pi a^2} \,dt = \frac{10^{-3.05} L_{\rm bol} t_{\rm sat}}{4\pi a^2},
\end{equation}

\noindent assuming that the star is currently in the unsaturated regime and that $L_{\rm bol}$ and $a$ do not depend on time.

This treatment also repeats the assumption from \citet{Lammer2009}; that is, that the ROSAT measurements provide an appropriate scaling for the XUV in general. To be more fully correct, the above expression should include a constant of proportionality; however, we will eventually normalize our $I_{\rm XUV}$ estimates with the fluence received at Earth, $I_{\rm XUV, \oplus}$. This normalization involves integrating Equation~\ref{eq:LX_G}, which is similarly affected by this assumption, and so such a constant will cancel out from our estimate of $\frac{I_{\rm XUV}}{I_{\rm XUV, \oplus}}$. For ease of reproducibility, we note that integration of Equation~\ref{eq:LX_G} from 0 to 4.6Gyr and division by $4\pi a_{\oplus}^2$ yields $I_{\rm XUV, \oplus}=2.0\times10^{18}$~erg cm$^{-2}$.

The above calculation re-raises the question of whether the ROSAT observations are truly a ``justified EUV proxy'' for M dwarfs. Observational evidence is limited since the EUV is inaccessible for most stars, but existing data are broadly consistent with this assumption: Figure 14 of \citet{Johnstone2021} finds a mass-independent scaling relation between X-ray and EUV flux that applies to both M dwarfs and Sun-like stars (albeit with substantial scatter), and Figure 15 of \citet{Pineda2021} finds that EUV emission of M dwarfs decreases by roughly two orders of magnitude between youth and field ages, similar to the findings of \citet{Wright2018} for X-rays. Nevertheless, \citet{King2021} find differing functional forms for $L_{\rm X}$/$L_{\rm bol}(t)$ and $L_{\rm EUV}$/$L_{\rm bol}(t)$ for Sun-like stars, and therefore it would be unsurprising if similar discrepancies are identified for M dwarfs with higher quality data. Uncertainty in the $L_{\rm EUV}$/$L_{\rm X}(t)$ profile for M dwarfs is thus a potential limitation of this work; however, we do not expect it to have a large impact on our results given that our calculations are dominated by the saturated regime where both profiles are flat, there is a two order-of-magnitude EUV decline for the unsaturated regime (and thus the assumption that the unsaturated EUV is negligible relative to the saturated EUV likely holds), and our model is insensitive to modest changes in the Rossby number of saturation, as discussed above.

\subsection{The mass-dependent saturation lifetime}
\label{sec:tsat}
How long do mid-to-late M dwarfs remain in the saturated regime? This is a challenging question to address directly, as it is notoriously difficult to measure ages of field M dwarfs \citep[e.g.,][]{Soderblom2010}. While open clusters could allow one to sidestep these difficulties, mid-to-late M dwarfs remain in the saturated regime at the age of Praesepe and the Hyades, and they are too faint to have had their rotation periods studied in older, more distant clusters \citep[e.g., Figure 15 of][]{Dungee2022}. Some works have attempted to answer this question using gyrochronology; for example, \citet{Engle2024} find an X-ray saturation lifetime of $\sim$2.3Gyr for mid-to-late M dwarfs based on ages they infer using the gyrochronological relation from \citet{Engle2023}. However, the use of gyrochronology may be problematic for fully convective M dwarfs: as we discuss in Section~\ref{sec:intro}, the gradual Skumanich-law spindown of these stars is dwarfed by the dramatic jump between rotation modes. These stars do not converge onto a tight mass-rotation sequence prior to making the jump, evidenced by the large dispersion we see in cluster rotation periods for mid-to-late M-dwarfs \citep[e.g., Figure 5 of][]{Pass2022}. For any individual mid-to-late M dwarf, rotation therefore does not provide a reliable estimate of age other than placing it within a broad category of ‘younger’ or ‘older’ (i.e., pre- or post-jump). In other words, mid-to-late M-dwarf rotation periods are not dominated by spindown as required by gyrochronology because they are greatly affected by inherent dispersion in initial rotation periods; a 0.2M$_\odot$ M dwarf with $P=5$ days could equally be a very young star with slow initial rotation, or it could be a few gigayears old and slowly spinning down from a faster initial rate \citep{Pass2022, Pass2023_active}.

As an alternative, we can adopt the statistical approach from \citet{Pass2024}. That work used the volume-complete sample of single, 0.1--0.3M$_\odot$ M dwarfs within 15pc \citep{Winters2021, Pass2023_inactive, Pass2023_active} to determine a mass-dependent equation for $t_{\rm sat}$, the lifetime of the saturated phase. In Gyr, this analysis yields
\begin{equation}
\label{eq:epoch}
t_{\rm sat}(M_*) = 5.9 - 15.4M_*/M_\odot.
\end{equation}

\noindent We note that this equation gives a consistent result with the \citet{Engle2024} saturation lifetime for a mass of 0.23M$_\odot$ ($\sim$M4), which is reasonable agreement given that their calculation averages over M2.5-6.5.

Equation~\ref{eq:epoch} has been fit only over the domain $0.1 \leq M_*/M_\odot\leq0.3$ and caution should be taken if extrapolating outside of these bounds. In particular, note that at the age of Praesepe (600--800Myr depending on the reference; see Table 1 of \citealt{Douglas2019} for review), partially convective M dwarfs with $M_* > 0.35M_\odot$ appear to have converged onto a slowly rotating sequence \citep[see Figure 5 of][]{Pass2022}, with Rossby numbers that place them in the unsaturated regime according to the relation of \citet{Wright2018}; further gradual spindown of early M dwarfs has been subsequently observed by the age of M67 \citep[4Gyr;][]{Dungee2022}. Such observations suggest that the rotational evolution of partially convective, early M dwarfs follows the same Skumanich-law spindown paradigm as K and G dwarfs. Our relation should therefore not be extrapolated beyond 0.35M$_\odot$: the bimodality in rotation/activity behavior is a trait specific to the fully convective M-dwarf population.

In the other direction, it is possible that ultracool M dwarfs have even longer lived activity than an extrapolation of our relation below 0.1M$_\odot$ would indicate; we lack the data to make a conclusive statement. Theoretically, such a behavior could result from weakened braking for ultracool dwarfs due to increased atmospheric neutrality at low temperatures \citep[e.g.,][]{Mohanty2002, Reiners2008}. Anecdotally, the 0.09M$_\odot$ TRAPPIST-1 remains in the saturated regime to the present day ($P_{\rm rot}$=$3.3$ days; \citealt{Vida2017}) despite reports that it is a transitional thin/thick disk star with an age estimate of 7.6$\pm$2.2Gyr \citep{Burgasser2017}. An extrapolation of our relation would suggest that TRAPPIST-1 transitions to the unsaturated regime at 4.5Gyr, marginally inconsistent with that age estimate (although accurate age estimates for M dwarfs are notoriously difficult). While it is possible that the saturation lifetime may deviate from linearity below 0.1M$_\odot$, there is also anecdotal evidence that ultracool dwarfs do spin down eventually at very advanced ages: Teegarden's Star is a member of the thick disk \citep{Fuhrmann2012}, suggesting an age of roughly 10Gyr, and it has been reported to be slowly rotating with a period of 100 days \citep{Terrien2022}.

\subsection{Sample selection and fundamental properties}
\label{sec:sample}
To select the planetary sample for our investigation, we query the Exoplanet Archive \citep{Akeson2013} for all confirmed transiting planets with $R_{\rm P} < 1.8R_\oplus$ out to 50pc that orbit stars with $M_* < 0.35M_\odot$, which we determine using Gaia parallaxes and the \citet{Benedict2016} $K$-band mass--luminosity relation to ensure consistency when comparing with Equation~\ref{eq:epoch}. These cuts yield a sample of 49 planets.

We estimate the bolometric luminosities of these stars using Gaia and 2MASS \citep{Skrutskie2006} photometry and distances from Gaia parallaxes \citep{Gaia2023}. We consider the bolometric corrections from both \cite{Pecaut2013} and \hbox{\cite{Mann2015}}, adopting the average of the luminosities calculated from the two methods. As these relations require $V$-band magnitudes, we estimate this value from ($G-K$) colors using an empirical relation (J.\ Winters, private communication) calibrated using the 15pc mid-to-late M-dwarf sample from \citet{Winters2021}:
\begin{equation}
V = K + 2.7638 - 0.5400(G-K) + 0.3094(G-K)^2.
\end{equation}

\noindent We elect to use this relation because it was calibrated specifically for mid-to-late M dwarfs, but if we instead employ the photometric transformation from Gaia DR3 \citep{Busso2022} that estimates $V$ from $G$ and \hbox{($G_{\rm BP}-G_{\rm RP}$)} color, our $L_{\rm bol}$ estimates change by no more than a few percent.

As we have restricted our sample to planets on the rocky side of the radius gap, we use the measured radius and the telluric mass--radius relation from \citet{Zeng2019} to estimate planetary masses, from which we proceed to calculate the planetary escape velocity, $v_{\rm esc}=\sqrt{\frac{M_P}{M_\oplus}\frac{R_\oplus}{R_P}}v_{\rm esc,\oplus}$.

We calculate the saturation lifetime for our sample stars using Equation~\ref{eq:epoch}, which includes extrapolating this relation into the 0.30--0.35M$_\odot$ mass regime. The only target in our sample with a mass below 0.1M$_\odot$ is TRAPPIST-1; as this star remains rapidly rotating at the present day, we adopt 7.6Gyr (its reported current age) as the endpoint for our integration, as motivated by the discussion in Section~\ref{sec:tsat}. If TRAPPIST-1 is truly younger than this estimate, the historic $I_{\rm XUV}$ flux received by its planets would be proportionally lower; its age uncertainty therefore has a significant impact on the predicted $I_{\rm XUV}$. For example, if TRAPPIST-1 is truly 4.5Gyr (its saturation lifetime as estimated by an extrapolation of Equation~\ref{eq:epoch} to lower masses), our $I_{\rm XUV}$ estimate would decrease by 40\%. Four other stars in our sample are also active and rapidly rotating at the present day; given the lack of reliable age estimates for M dwarfs, we use their saturation lifetimes as the endpoint of integration and present their $I_{\rm XUV}$ estimates with the caveat that they are upper limits.

\subsection{Influence of PMS phase and energetic flares}
\label{sec:higher}
We are now equipped to apply Equation~\ref{eq:our_Ixuv} to the planetary sample defined in the previous section, thereby determining the historic XUV fluence received by these planets and informing us of their capacity for atmosphere retention. However, there are additional effects that it may be prudent to consider, particularly those phenomena that have a more significant impact on mid-to-late M dwarfs than Sun-like stars. Here we will briefly discuss and model the effects of pre-main-sequence overluminosity and energetic flares.

\subsubsection{Pre-main-sequence overluminosity}
\label{sec:PMS}
In writing Equation~\ref{eq:our_Ixuv}, we assumed that $L_{\rm bol}$ does not depend on time. Such an assumption does not acknowledge the star's heightened luminosity during the pre-main-sequence (PMS) phase, which can result in hundreds of millions of years of overluminosity for very low mass stars.

To investigate how the inclusion of the PMS phase would affect our estimates, we consider solar-metallicity MIST evolutionary tracks \citep{Dotter2016, Choi2016}. We make the assumption that $\big(\frac{L_{\rm X}}{L_{\rm bol}}\big)_{\rm sat} = 10^{-3.05}$ still holds during the PMS phase. Our equation thus becomes
\begin{equation}\label{eq:our_Ixuv_pms}
I_{\rm XUV} = \frac{10^{-3.05} (f_{\rm L} L_{\rm bol}) t_{\rm sat}}{4\pi a^2},
\end{equation}

\noindent where ($f_{\rm L}L_{\rm bol}$) is the average luminosity of the star over the entire saturated phase, including the PMS period.

To determine $f_{\rm L}$, we truncate the MIST evolutionary tracks at $t_{\rm sat}$, normalize $L(t)$ by its value at 1Gyr (which approximates our measured $L_{\rm bol}$, the bolometric luminosity on the main sequence), and calculate the mean value of $L(t)/L_{\rm bol}$ from the earliest time step to $t_{\rm sat}$.

\begin{table}[t]
\centering
\begin{tabular}{llllll}
\toprule
{$M_*$} & {$L_{\rm bol}$} & {$t_{\rm sat}$} & {$f_{\rm L}$} & {$I_{\rm XUV, base}$} & {$I_{\rm XUV, corr}$} \\
{[$M_\odot$]} & {[$L_\odot$]} & {[Gyr]} & {} & {[${I_{\rm XUV, \oplus}}$]} & {[${I_{\rm XUV, \oplus}}$]} \\
\midrule
0.10 & 0.00086 & 4.4  & 1.19 & $\frac{0.073}{(a[\rm{au}])^2}$ & $\frac{0.11}{(a[\rm{au}])^2}$ \\
0.12 & 0.0018  & 4.1  & 1.17 & $\frac{0.14}{(a[\rm{au}])^2}$  & $\frac{0.21}{(a[\rm{au}])^2}$ \\
0.15 & 0.0031  & 3.6  & 1.13 & $\frac{0.22}{(a[\rm{au}])^2}$  & $\frac{0.31}{(a[\rm{au}])^2}$ \\
0.20 & 0.0053  & 2.8  & 1.13 & $\frac{0.29}{(a[\rm{au}])^2}$  & $\frac{0.41}{(a[\rm{au}])^2}$ \\    
0.25 & 0.0075  & 2.1  & 1.16 & $\frac{0.30}{(a[\rm{au}])^2}$  & $\frac{0.43}{(a[\rm{au}])^2}$ \\
0.30 & 0.010   & 1.3  & 1.22 & $\frac{0.25}{(a[\rm{au}])^2}$  & $\frac{0.38}{(a[\rm{au}])^2}$ \\
\bottomrule
\end{tabular}
\caption{Grid of tabulated parameters for mid-to-late M dwarfs. To obtain average values of $L_{\rm bol}$ as a function of mass, we fit a fourth-order polynomial to the $M_*$ vs $L_{\rm bol}$ distribution for the 200 single, inactive, 0.1--0.3M$_\odot$ M dwarfs within 15pc from \citet{Pass2023_inactive}. We estimate the PMS correction, $f_{\rm L}$, from MIST models, as described in Section~\ref{sec:PMS}. The two $I_{\rm XUV}$ columns follow from Equations~\ref{eq:our_Ixuv} and \ref{eq:our_Ixuv_corr}, indicating our estimate without and with the PMS/flare corrections.}\label{tab:grid}
\end{table}

\begin{figure*}[t]
    \centering
    \includegraphics[width=\textwidth]{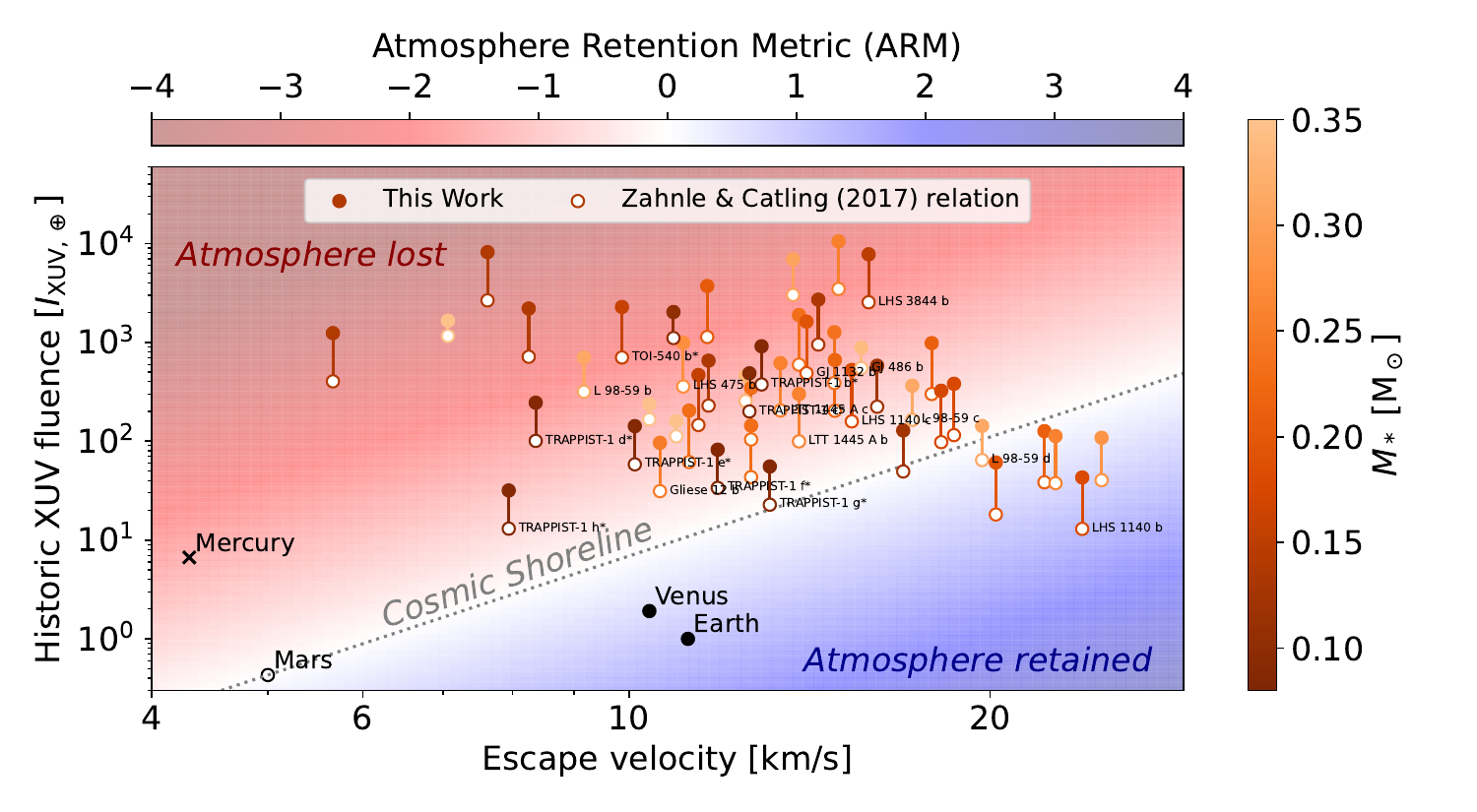}
    \caption{The sample of known transiting terrestrial planets that orbit mid-to-late M dwarfs within 50pc. Our estimates are noted by solid circles, including PMS and flare corrections. For comparison, the predictions from the \citet{Zahnle2017} scaling law are shown as open circles. We find that our revised estimates result in historic XUV fluences that are 2.1 times the scaling relation on average, or 3.1 times after considering the PMS and flare corrections. The dashed line indicates the cosmic shoreline as defined in \citet{Zahnle2017}: $I_{\rm XUV} \propto v_{\rm esc}^4$, with the constant of proportionality defined such that Mars sits on the shoreline. For $I_{\rm XUV}$ in Earth units and $v_{\rm esc}$ in kms$^{-1}$, this equation can be written as $I_{\rm XUV} = 6.9 \times 10^{-4} v_{\rm esc}^4$. The underlying heatmap is colored by our Atmosphere Retention Metric (ARM), indicating the log-space distance from the cosmic shoreline. For legibility, only stars within 15pc are labeled by name; however, all values are given in Table~\ref{tab:planets}. Stars that are still rapidly rotating at the present day are noted with an asterisk; our estimates may be inaccurate in these cases, as the correct answer will be dependent on the age of the star and reliable age estimates for M dwarfs are generally unavailable.}
    \label{fig:cosmic}
\end{figure*}

\begin{table*}[!p]
\centering\scriptsize\hspace*{-2.1cm}
\begin{tabular}{llllllll}
\toprule
{Planet}       & {$M_*$}   &  {$L_{\rm bol}$}      & {$a$}      & {$v_{\rm esc}$} & {$I_{\rm XUV, base}$}            & {$I_{\rm XUV, corr}$}  & {ARM}         \\
             & {[$M_\odot$]} & {[$L_{\odot}$]}  & {[au]}     & {[kms$^{-1}$]} & {[${I_{\rm XUV, \oplus}}$]} & {[${I_{\rm XUV, \oplus}}$]} & {}\\
\midrule
Kepler-42 d  & 0.14      & 0.0028    & 0.015  & 5.7  & 8.7E+02                 & 1.2E+03 & $-$3.24                 \\
LHS 1678 b   & 0.35      & 0.014     & 0.012  & 7.1  & 9.3E+02                 & 1.7E+03  & $-$2.98               \\
Kepler-42 c  & 0.14      & 0.0028    & 0.0060 & 7.6  & 5.7E+03                 & 8.2E+03  & $-$3.55               \\
TRAPPIST-1 h$^{**}$ & 0.088     & 0.00054   & 0.061  & 7.9  & 2.1E+01                 & 3.2E+01 & $-$1.06                \\
Kepler-42 b  & 0.14      & 0.0028    & 0.012  & 8.3  & 1.5E+03                 & 2.2E+03 & $-$2.84              \\
TRAPPIST-1 d$^{**}$ & 0.088     & 0.00054   & 0.022  & 8.4  & 1.6E+02                 & 2.5E+02 & $-$1.86                \\
L 98-59 b    & 0.31      & 0.011     & 0.023  & 9.2  & 4.5E+02                 & 7.0E+02 & $-$2.16              \\
TOI-540 b$^{*}$    & 0.16      & 0.0036    & 0.012  & 9.9  & 1.6E+03                 & 2.3E+03 & $-$2.54                 \\
TRAPPIST-1 e$^{**}$ & 0.088     & 0.00054   & 0.029  & 10.1 & 9.4E+01                 & 1.4E+02  & $-$1.29            \\
LHS 1678 c   & 0.35      & 0.014     & 0.033  & 10.4 & 1.3E+02                 & 2.4E+02  & $-$1.47               \\
Gliese 12 b  & 0.25      & 0.0075    & 0.067  & 10.6 & 6.7E+01                 & 9.6E+01 & $-$1.04                \\
SPECULOOS-3 b$^{*}$ & 0.10 & 0.00087 & 0.0073 & 10.9 & 1.4E+03 & 2.0E+03 & $-$2.32 \\
LHS 1678 d   & 0.35      & 0.014     & 0.040  & 11.0 & 9.0E+01                 & 1.6E+02  & $-$1.21               \\
LHS 475 b    & 0.27      & 0.0085    & 0.020  & 11.1 & 6.7E+02                 & 9.9E+02  & $-$1.98            \\
K2-239 c     & 0.23      & 0.0069    & 0.047  & 11.2 & 1.4E+02                 & 2.0E+02  & $-$1.27               \\
K2-415 b$^{*}$     & 0.16      & 0.0035    & 0.027  & 11.4 & 3.3E+02                 & 4.7E+02   & $-$1.60              \\
TOI-6008 b   & 0.20      & 0.0052    & 0.010  & 11.6 & 2.6E+03                 & 3.7E+03  & $-$2.47               \\
LP 791-18 d  & 0.14      & 0.0024    & 0.020  & 11.7 & 4.5E+02                 & 6.5E+02   & $-$1.71              \\
TRAPPIST-1 f$^{**}$ & 0.088     & 0.00054   & 0.038  & 11.9 & 5.4E+01                 & 8.2E+01  & $-$0.78               \\
GJ 3929 b    & 0.33      & 0.012     & 0.026  & 12.5 & 2.8E+02                 & 4.6E+02  & $-$1.44               \\
TRAPPIST-1 c$^{**}$ & 0.088     & 0.00054   & 0.016  & 12.6 & 3.2E+02                 & 4.9E+02  & $-$1.45              \\
K2-239 b     & 0.23      & 0.0069    & 0.036  & 12.6 & 2.4E+02                 & 3.5E+02 & $-$1.29                \\
K2-239 d     & 0.23      & 0.0069    & 0.056  & 12.6 & 1.0E+02                 & 1.4E+02 & $-$0.91                \\
TRAPPIST-1 b$^{**}$ & 0.088     & 0.00054   & 0.011  & 12.9 & 6.1E+02                 & 9.1E+02 & $-$1.68                \\
TRAPPIST-1 g$^{**}$ & 0.088     & 0.00054   & 0.046  & 13.1 & 3.7E+01                 & 5.5E+01  & $-$0.44               \\
LTT 1445 A c & 0.26      & 0.0080    & 0.027  & 13.4 & 4.2E+02                 & 6.2E+02  & $-$1.45               \\
TOI-1442 b   & 0.31      & 0.010     & 0.0073 & 13.7 & 4.5E+03                 & 6.9E+03 & $-$2.45                \\
LTT 1445 A b & 0.26      & 0.0080    & 0.038  & 13.9 & 2.1E+02                 & 3.0E+02 & $-$1.07                \\
TOI-6086 b   & 0.23      & 0.0065    & 0.015  & 13.9 & 1.3E+03                 & 1.9E+03 & $-$1.87                \\
GJ 1132 b    & 0.19      & 0.0049    & 0.016  & 14.1 & 1.2E+03                 & 1.6E+03 & $-$1.78                \\
LP 791-18 b  & 0.14      & 0.0024    & 0.0097 & 14.4 & 1.9E+03                 & 2.7E+03 & $-$1.96                \\
LHS 1478 b   & 0.24      & 0.0071    & 0.019  & 14.8 & 8.8E+02                 & 1.3E+03 & $-$1.58                \\
TOI-2096 b   & 0.22      & 0.0060    & 0.025  & 14.8 & 4.6E+02                 & 6.6E+02 & $-$1.30                \\
TOI-2445 b   & 0.25      & 0.0076    & 0.0064 & 15.0 & 7.3E+03                 & 1.1E+04  & $-$2.49               \\
LHS 1140 c   & 0.18      & 0.0044    & 0.027  & 15.3 & 3.7E+02                 & 5.3E+02  & $-$1.14               \\
GJ 486 b     & 0.34      & 0.012     & 0.018  & 15.6 & 5.2E+02                 & 8.9E+02  & $-$1.34               \\
LHS 3844 b   & 0.15      & 0.0031    & 0.0062 & 15.8 & 5.5E+03                 & 7.8E+03  & $-$2.25               \\
LP 890-9 b   & 0.12      & 0.0018    & 0.019  & 16.1 & 4.0E+02                 & 5.8E+02  & $-$1.10               \\
LP 890-9 c   & 0.12      & 0.0018    & 0.040  & 16.9 & 8.8E+01                 & 1.3E+02  & $-$0.36               \\
L 98-59 c    & 0.31      & 0.011     & 0.032  & 17.2 & 2.3E+02                 & 3.6E+02  & $-$0.78               \\
TOI-771 b    & 0.21      & 0.0056    & 0.020  & 17.9 & 6.9E+02                 & 9.8E+02  & $-$1.15               \\
TOI-237 b    & 0.17      & 0.0042    & 0.034  & 18.2 & 2.3E+02                 & 3.2E+02  & $-$0.63               \\
TOI-1680 b   & 0.18      & 0.0043    & 0.031  & 18.7 & 2.7E+02                 & 3.8E+02  & $-$0.66               \\
L 98-59 d    & 0.31      & 0.011     & 0.051  & 19.7 & 9.1E+01                 & 1.4E+02  & $-$0.14               \\
TOI-715 b    & 0.20      & 0.0052    & 0.082  & 20.2 & 4.3E+01                 & 6.1E+01  & +0.28               \\
TOI-6002 b   & 0.21      & 0.0057    & 0.057  & 22.2 & 8.9E+01                 & 1.3E+02  & +0.12               \\
TOI-1452 b   & 0.26      & 0.0078    & 0.062  & 22.7 & 7.7E+01                 & 1.1E+02  & +0.21               \\
LHS 1140 b   & 0.18      & 0.0044    & 0.094  & 23.9 & 3.0E+01                 & 4.3E+01  & +0.72               \\
TOI-5713 b$^{*}$   & 0.28      & 0.0090    & 0.061  & 24.8 & 7.3E+01                 & 1.1E+02 & +0.38  \\
\bottomrule
\end{tabular}
\caption{Our estimated historic XUV fluences for the known $R_{\rm P} < 1.8R_\oplus$ planets orbiting mid-to-late M dwarfs within 50pc; these planets are plotted in Figure~\ref{fig:cosmic}. Masses, bolometric luminosities, and escape velocities are calculated following Section~\ref{sec:sample}. Semimajor axes are calculated from the tabulated $M_*$ and the period reported in the Exoplanet Archive \citep{Akeson2013}. The two $I_{\rm XUV}$ columns follow from Equations~\ref{eq:our_Ixuv} and \ref{eq:our_Ixuv_corr}, indicating our estimates without and with the PMS/flare corrections. The Atmosphere Retention Metric (ARM; Equation~\ref{eq:arm}) indicates the distance between our $I_{\rm XUV, corr}$ estimate and the cosmic shoreline, with negative values meaning that the planet is above the shoreline and total atmospheric loss is probable.\\ \\ \footnotesize{$^*$ These planets' discovery papers report that the star is currently rapidly rotating \citep{Ment2021, Hirano2023, Ghachoui2024, Gillon2024}. For stars that are still in the saturated regime, we should perform the integral in Equation~\ref{eq:our_Ixuv} only to the current age of the star, not to $t_{\rm sat}$; however, a reliable age estimate is not available. Our $I_{\rm XUV}$ estimates are thus an upper limit.}
\\ \footnotesize{$^{**}$ TRAPPIST-1 is also reported to be in the saturated regime at the present day; see discussion in Section~\ref{sec:tsat}. To calculate our $I_{\rm XUV}$} estimates, we have performed the integral up to 7.6Gyr, the age estimate from \citet{Burgasser2017}.}\label{tab:planets}
\end{table*}

For 0.1--0.3M$_\odot$ M dwarfs, we find that the inclusion of the PMS phase modestly increases the $I_{\rm XUV}$ fluence by 13--22\%. Specific values are tabulated in Table~\ref{tab:grid}. Note that the smallest correction of 13\% is for an intermediate mass of 0.2M$_\odot$, with the correction rising to 19\% at 0.1M$_\odot$ and 22\% at 0.3M$_\odot$. This is because there are two competing effects at play: as mass decreases, the PMS phase is longer and contributes more overluminosity in an absolute sense; however, the saturation lifetime is also much longer, and so the PMS phase nonetheless represents a smaller fraction of that total.

\subsubsection{Energetic flares}

\citet{Medina2020, Medina2022_flare} observe that the flare rate of mid-to-late M dwarfs follows a similar saturated/unsaturated relationship as $\frac{L_{\rm X}}{L_{\rm bol}}$, with log($R_{31.5}$), the number of flares per day with energies above $3.16\times10^{31}$ erg in the TESS bandpass, saturating at $-1.32 \pm 0.06$. For mid-to-late M dwarfs in the unsaturated regime, this rate is decreased by 2--12 orders of magnitude. It is therefore reasonable to treat the flaring behavior of these stars in a similar manner to the X-ray luminosity: at times younger than $t_{\rm sat}$, the planet received a contribution from energetic flares that is constant in time; at later times, the contribution can be considered negligible in comparison (although note the caveat that while they have a negligible contribution to the overall $I_{\rm XUV}$ budget, flares capable of driving hydrodynamic escape still occur during the unsaturated phase and may therefore have a significant impact on the retention of secondary atmospheres; e.g., \citealt{France2020}).

To quantify the energetic flare contribution during the saturated phase, we follow the treatment in \citet{Ribas2016}, which draws from the analysis of CN Leo in \citet{Audard2000}. CN Leo is a saturated-regime, low-mass M dwarf, and thus its flaring behaviors are representative of the saturated phase that we seek to model. By analyzing the CN Leo flare distribution, \citet{Ribas2016} conclude that highly energetic flares increase the average X-ray dose by about 25\% over its typically observed value. We thus modify our equation for $I_{\rm XUV}$ by an additional factor $f_{\rm flare}$,
\begin{equation}\label{eq:our_Ixuv_corr}
I_{\rm XUV} = \frac{(f_{\rm flare}10^{-3.05}) (f_{\rm L} L_{\rm bol}) t_{\rm sat}}{4\pi a^2},
\end{equation}

\noindent where $f_{\rm flare}=1.25$.

\section{Results}
\label{sec:results}
\subsection{Comparison with \citet{Zahnle2017}}
In Figure~\ref{fig:cosmic}, we plot the historic XUV fluence that we calculate for each of the mid-to-late M dwarf terrestrial planets selected in Section~\ref{sec:sample}, with these measurements tabulated in Table~\ref{tab:planets}. This includes both the results of our Equation~\ref{eq:our_Ixuv} calculation, as well as the higher order corrections described in Section~\ref{sec:higher} and Equation~\ref{eq:our_Ixuv_corr}. We also compare these calculations to the \citet{Zahnle2017} scaling law (Equation~\ref{eq:Ixuv}).

\begin{figure}[t]
    \centering
    \includegraphics[width=\columnwidth]{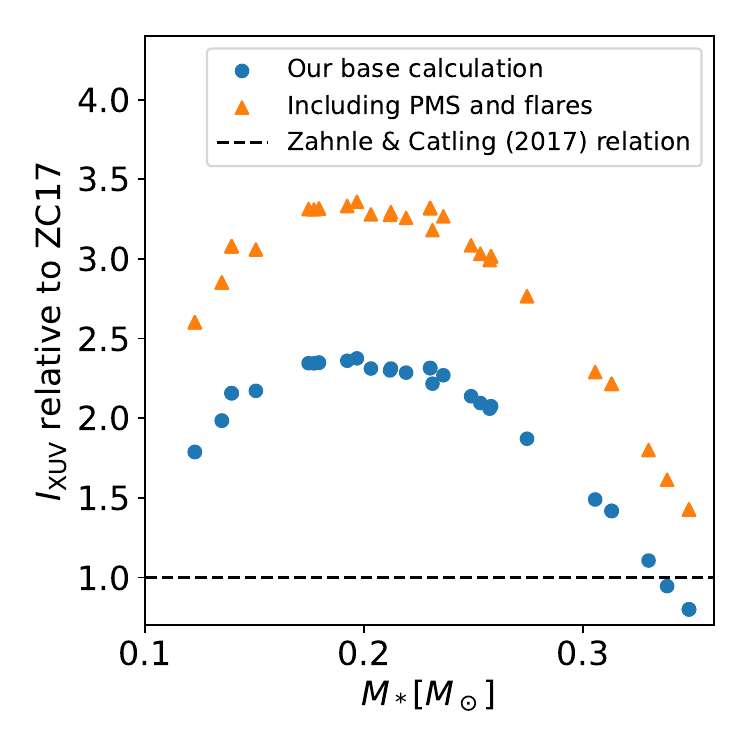}
    \caption{The ratio of $I_{\rm XUV}$ as calculated in this work relative to that calculated using the \citet{Zahnle2017} scaling law (Equation~\ref{eq:Ixuv}) for the known $R_{\rm P} < 1.8R_\oplus$ planets orbiting inactive mid-to-late M dwarfs within 50pc. Circles indicate the results of our base equation (Equation~\ref{eq:our_Ixuv}) while triangles include corrections for pre-main-sequence overluminosity and energetic flares (Equation~\ref{eq:our_Ixuv_corr}).}
    \label{fig:ratio}
\end{figure}

We highlight the differences between these relations in Figure~\ref{fig:ratio}, where we plot the ratio between our calculations and the \citet{Zahnle2017} relation as a function of mass. For our base relation, we calculate a historic XUV fluence 2.1 times that of the \citet{Zahnle2017} scaling law on average, finding the largest discrepancies near 0.2M$_\odot$. At these intermediate masses, our base calculation yields 2.3 times the XUV fluence, and 3.3 times when the flare and PMS corrections are included. At the highest masses, our base relation agrees with the \citet{Zahnle2017} scaling law within 20\%; this is perhaps unsurprising given that the 0.6Gyr timescale used in Equation~\ref{eq:LX_M} is a reasonable assumption for the saturation lifetime of the most massive fully convective M dwarfs.

\subsection{The cosmic shoreline}
We plot the cosmic shoreline in Figure~\ref{fig:cosmic} following the definition from \citet{Zahnle2017}: this line indicates the relationship $I_{\rm XUV} \propto v_{\rm esc}^4$, with the constant of proportionality set such that Mars sits on the shoreline. To quantify the position of a planet relative to this shoreline, we define the Atmosphere Retention Metric (ARM) as
\begin{equation}
\label{eq:arm}
\mathrm{ARM} \equiv 4\log_{10}(v_{\rm esc}) - \log_{10}(I_{\rm XUV}) - 3.16,
\end{equation}

\noindent with $I_{\rm XUV}$ in Earth units and $v_{\rm esc}$ in kms$^{-1}$. The ARM is proportional to the log-space distance between a point ($v_{\rm esc}$, $I_{\rm XUV}$) and the cosmic shoreline, with a negative value indicating that the planet is above the shoreline and complete atmospheric loss is probable.

Given the $I_{\rm XUV}$ values estimated in this work, few nearby mid-to-late M-dwarf terrestrial planets fall below the cosmic shoreline. Invariably these best candidates for atmosphere retention are the largest, with those planets with $v_{\rm esc} \geq 20$kms$^{-1}$ all falling in the favorable regime. All of the smaller planets known to exist within 50pc fall above the $I_{\rm XUV} \propto v_{\rm esc}^4$ shoreline in light of our revised estimates of $I_{\rm XUV}$. Super Earths may therefore be our only hope for the detection of a terrestrial planet atmosphere through thermal emission measurements with JWST, at least within the cosmic shoreline paradigm of \citet{Zahnle2017}. That said, this paradigm is a simplification of the complex processes that govern escape and more sophisticated treatments may be able to save some planets from an airless fate \citep[e.g.,][]{Nakayama2022, Chatterjee2024}, although a detailed discussion of such processes is beyond the scope of this work.

In addition, there is a possibility that some of these large worlds may not truly be terrestrial: indeed, in each of the four cases for which there is a mass measurement reported in the Exoplanet Archive for a planet in our sample with $R_{\rm P} > 1.3$R$_{\oplus}$ (L 98-59 c and d, TOI-1452 b, and LHS 1140 b; \citealt{Demangeon2021, Cadieux2022, Cadieux2024}), the measurement is smaller than that predicted by the telluric composition model (although in the cases of L 98-59 c and LHS 1140 b, other works do find a terrestrial composition; c.f. \citealt{Cloutier2019, Ment2019}). Such discrepancies between RV mass measurements determined by different studies are not uncommon; in a meta analysis of 22 different methods tested on the same RV dataset, \citet{Zhao2022} noted ``a concerning lack of agreement between the RVs returned by different methods,'' generally resulting from how much signal is subsumed into stellar activity. The composition of these worlds is therefore not currently conclusive. If the low mass measurements prove to be accurate, the escape velocities of these planets would be lower than our estimates from the telluric model, pushing them closer to the cosmic shoreline (e.g., a 20\% change in $v_{\rm esc}$ corresponds to $\pm$0.3 for the ARM).

Some planets in our sample have already been determined to be airless based on secondary eclipse and/or phase curve measurements. These planets are LHS~3844~b \citep{Kreidberg2019}, TRAPPIST-1~b \citep{Greene2023}, TRAPPIST-1~c \citep{Zieba2023}, LTT~1445~A~b \citep{Wachiraphan2024}, GJ~1132~b \citep{Xue2024}, and Gl~486~b \citep{WeinerMansfield2024}. These planets all fall above the cosmic shoreline using either the \citet{Zahnle2017} relation or the calculations presented in this work, and so their airlessness does not provide particularly tight constraints on the cosmic shoreline. \citet{August2025} report a tentative detection of an atmosphere for LHS~1478~b, which falls above the cosmic shoreline when considering either framework and hence would be an intriguing constraint on atmosphere retention; however, they note that the two observations from their program do not yield consistent results. For this reason, other works such as \citet{ParkCoy2024} have neglected this measurement from their population-level analysis of rocky planets with emission data. 

Note that we only consider fully convective M dwarfs with masses $\leq0.35M_\odot$ in this work, as our key insight is that the active lifetimes of these stars have been underestimated by the canonical $I_{\rm XUV}$ scaling relation. Some rocky worlds orbiting earlier M dwarfs have also been studied with emission methods, but they are not members of our sample \citep{Crossfield2022, Zhang2024, Luque2024, MeierValdes2025}.

\subsection{Notes on uncertainties}
How might uncertainties in various parameters affect the $I_{\rm XUV}$ dose received by a planet? Our equation for $I_{\rm XUV}$ scales linearly with the saturation lifetime, $t_{\rm sat}$, and the saturation level, $\big(\frac{L_{\rm X}}{L_{\rm bol}}\big)_{\rm sat}$, making these key parameters to consider to understand potential errors.

The analysis of M--M wide binary pairs in \citet{Pass2024} places an upper limit on the intrinsic scatter in $t_{\rm sat}$ of 25\%, indicating that star-to-star variations in saturation lifetime should generally have only a modest effect on XUV fluence, perhaps comparable in amplitude to the inclusion/exclusion of the energetic flare correction. This corresponds to an error of $\pm$0.1 in the ARM. There may, however, be some exceptions to this rule: \citet{Pass2022} identified a handful of 0.2--0.3M$_\odot$ M dwarfs that appeared to have made the jump at much younger ages than Equation~\ref{eq:epoch} would indicate, systems which may represent special initial conditions (such as their high-energy birth environment and initial rotation rate) that could allow for a shorter saturation lifetime and an increased likelihood of planetary atmosphere retention. While such systems would be intriguing for atmospheric characterization, we do not currently have a method to identify which systems that are in the unsaturated regime today may have experienced such an anomalous XUV history, exempting special cases such as those mentioned in \citet{Pass2022} where we have age estimates from chance circumstances, such as cluster membership or a wide binary companion.

Another potential source of uncertainty is our limited knowledge of the mechanism surrounding the saturation level, $\big(\frac{L_{\rm X}}{L_{\rm bol}}\big)_{\rm sat}$. While the nominal error in $\big(\frac{L_{\rm X}}{L_{\rm bol}}\big)_{\rm sat}$ reported by \citet{Wright2018} is small, corresponding to a change in $I_{\rm XUV}$ of 12\%, there is nonetheless substantial dispersion within the saturated regime; \citet{Wright2011} note an rms residual of roughly 0.3 dex after fitting Equation~\ref{eq:saturate}, which would correspond to roughly a factor of 2 in star-to-star variation in $\frac{L_{\rm X}}{L_{\rm bol}}$, and hence potentially a factor of 2 variation in $I_{\rm XUV}$ (and an error of $\pm$0.3 in the ARM). \citet{Wright2011} attribute the scatter to a number of factors, including the variability of X-ray luminosity over the stellar cycle and uncertainties in stellar parameters. They also note: ``X-ray saturation [may] not be an actual saturation in any sense, but a completely different magnetic dynamo configuration with completely different dependencies.'' It is therefore unclear if there exists true star-to-star variation in the saturation level. Future advances in our theoretical understanding of the magnetic dynamo may thus lead to a reduction in this source of uncertainty.

Less quantifiable are the uncertainties intrinsic to the simplicity of the cosmic shoreline paradigm; as previously mentioned, detailed modeling with more sophisticated physics may change the results relative to this simplistic picture, and there are uncertainties involved in treating X-ray observations as a proxy for the EUV. The choice of proportionality constant for the $I_{\rm XUV} \propto v_{\rm esc}^4$ relation (i.e., whether the shoreline is drawn directly through Mars, or slightly above or below it) can also lead to modest changes in ARM. Overall, the ARM should be interpreted as a prioritization metric: a planet with a higher ARM is more likely to be able to retain an atmosphere than a planet with a lower ARM, but we are not intending to suggest that planets with negative ARMs should be discounted entirely from further study.

\section{Conclusions}
In this paper, we have highlighted recent developments in our understanding of mid-to-late M-dwarf rotation/activity evolution and applied these insights to the question of the cosmic shoreline. In particular, we note that mid-to-late M dwarfs remain in the saturated regime for billions of years longer than early M dwarfs and Sun-like stars, and spend little of their lives with intermediate activity levels. We present a model of historic XUV fluence that accounts for this evolutionary history. For the average mid-to-late M-dwarf planet, our revised calculation of $I_{\rm XUV}$ doubles the received XUV dose as compared to the canonical \citet{Zahnle2017} scaling relation, or triples it if also considering the extended pre-main-sequence phase and high-energy flares (Figure~\ref{fig:ratio}), causing the cosmic shoreline to recede from these worlds (Figure~\ref{fig:cosmic}).

For the reader interested in calculating the historic XUV fluence received by a mid-to-late M-dwarf planet, we recommend using our Equation~\ref{eq:our_Ixuv_corr}, which accounts for the bimodal rotation/activity evolution of these stars and includes pre-main-sequence and energetic flare corrections. For convenience, we also provide a pre-computed grid of $I_{\rm XUV}$ estimates as a function of $M_{*}$ and $a$ in Table~\ref{tab:grid}, and report the estimates for the 49 planets investigated in this analysis in Table~\ref{tab:planets}. Furthermore, we provide an Atmosphere Retention Metric (ARM; Equation~\ref{eq:arm}) to calculate the distance between a planet and the cosmic shoreline.

Within the cosmic shoreline paradigm of \citet{Zahnle2017}, the vast majority of known terrestrial planets of mid-to-late M dwarfs are unlikely to be able to retain atmospheres. The most likely exceptions are the planets nearest to the radius gap, assuming they indeed prove to be terrestrial. These super Earths may thus represent the most compelling targets for JWST in its quest to discover the atmosphere of a rocky exoplanet through emission methods.

\section*{Acknowledgments}

E.P.\ is supported by a Juan Carlos Torres Postdoctoral Fellowship at the Massachusetts Institute of Technology and A.V.\ is supported in part by a Sloan Research Fellowship.

This research has made use of:

\begin{itemize}[leftmargin=*]
\item The NASA Exoplanet Archive, which is operated by the California Institute of Technology, under contract with the National Aeronautics and Space Administration under the Exoplanet Exploration Program.
\item The European Space Agency (ESA) mission Gaia (\url{https://www.cosmos.esa.int/gaia}), processed by the Gaia Data Processing and Analysis Consortium (DPAC, \url{https://www.cosmos.esa.int/web/gaia/dpac/consortium}). Funding for the DPAC has been provided by national institutions, in particular the institutions participating in the Gaia Multilateral Agreement.
\item The Two Micron All Sky Survey, which is a joint project of the University of Massachusetts and the Infrared Processing and Analysis Center/California Institute of Technology, funded by the National Aeronautics and Space Administration and the National Science Foundation.
\end{itemize}
%

\software{\texttt{Matplotlib} \citep{Hunter2007}, \texttt{NumPy} \citep{Harris2020}, \texttt{pandas} \citep{Reback2021}, \texttt{SciPy} \citep{Scipy2020}}


\bibliography{cosmic-shoreline}{}

\begin{thebibliography}{87}
\expandafter\ifx\csname natexlab\endcsname\relax\def\natexlab#1{#1}\fi

\bibitem[{{Akeson} {et~al.}(2013){Akeson}, {Chen}, {Ciardi}, {Crane}, {Good}, {Harbut}, {Jackson}, {Kane}, {Laity}, {Leifer}, {Lynn}, {McElroy}, {Papin}, {Plavchan}, {Ram{\'\i}rez}, {Rey}, {von Braun}, {Wittman}, {Abajian}, {Ali}, {Beichman}, {Beekley}, {Berriman}, {Berukoff}, {Bryden}, {Chan}, {Groom}, {Lau}, {Payne}, {Regelson}, {Saucedo}, {Schmitz}, {Stauffer}, {Wyatt}, \& {Zhang}}]{Akeson2013}
{Akeson}, R.~L., {Chen}, X., {Ciardi}, D., {et~al.} 2013, \href{http://dx.doi.org/10.1086/672273}{\color{magenta}\pasp}, \href{https://ui.adsabs.harvard.edu/abs/2013PASP..125..989A}{125, 989}

\bibitem[{{Audard} {et~al.}(2000){Audard}, {G{\"u}del}, {Drake}, \& {Kashyap}}]{Audard2000}
{Audard}, M., {G{\"u}del}, M., {Drake}, J.~J., \& {Kashyap}, V.~L. 2000, \href{http://dx.doi.org/10.1086/309426}{\color{magenta}\apj}, \href{https://ui.adsabs.harvard.edu/abs/2000ApJ...541..396A}{541, 396}

\bibitem[{{August} {et~al.}(2025){August}, {Buchhave}, {Diamond-Lowe}, {Mendon{\c{c}}a}, {Gressier}, {Rathcke}, {Allen}, {Fortune}, {Jones}, {Meier Vald{\'e}s}, {Demory}, {Espinoza}, {Fisher}, {Gibson}, {Heng}, {Hoeijmakers}, {Hooton}, {Kitzmann}, {Prinoth}, {Eastman}, \& {Barnes}}]{August2025}
{August}, P.~C., {Buchhave}, L.~A., {Diamond-Lowe}, H., {et~al.} 2025, \href{http://dx.doi.org/10.1051/0004-6361/202452611}{\color{magenta}\aap}, \href{https://ui.adsabs.harvard.edu/abs/2025A&A...695A.171A}{695, A171}

\bibitem[{Benedict {et~al.}(2016)Benedict, Henry, Franz, McArthur, Wasserman, Jao, Cargile, Dieterich, Bradley, Nelan, \& Whipple}]{Benedict2016}
Benedict, G.~F., Henry, T.~J., Franz, O.~G., {et~al.} 2016, \href{http://dx.doi.org/10.3847/0004-6256/152/5/141}{\color{magenta}AJ}, \href{https://ui.adsabs.harvard.edu/abs/2016AJ....152..141B}{152, 141}

\bibitem[{{Burgasser} \& {Mamajek}(2017)}]{Burgasser2017}
{Burgasser}, A.~J. \& {Mamajek}, E.~E. 2017, \href{http://dx.doi.org/10.3847/1538-4357/aa7fea}{\color{magenta}\apj}, \href{https://ui.adsabs.harvard.edu/abs/2017ApJ...845..110B}{845, 110}

\bibitem[{{Busso} {et~al.}(2022){Busso}, {Cacciari}, {Bellazzini}, {Carrasco}, {De Angeli}, {Evans}, {Fabricius}, {Montegriffo}, {Pancino}, {Rainer}, \& {Sanna}}]{Busso2022}
{Busso}, G., {Cacciari}, C., {Bellazzini}, M., {et~al.} 2022, {Gaia DR3 documentation Chapter 5: Photometric data}

\bibitem[{{Cadieux} {et~al.}(2022){Cadieux}, {Doyon}, {Plotnykov}, {H{\'e}brard}, {Jahandar}, {Artigau}, {Valencia}, {Cook}, {Martioli}, {Vandal}, {Donati}, {Cloutier}, {Narita}, {Fukui}, {Hirano}, {Bouchy}, {Cowan}, {Gonzales}, {Ciardi}, {Stassun}, {Arnold}, {Benneke}, {Boisse}, {Bonfils}, {Carmona}, {Cort{\'e}s-Zuleta}, {Delfosse}, {Forveille}, {Fouqu{\'e}}, {Gomes da Silva}, {Jenkins}, {Kiefer}, {K{\'o}sp{\'a}l}, {Lafreni{\`e}re}, {Martins}, {Moutou}, {do Nascimento}, {Ould-Elhkim}, {Pelletier}, {Twicken}, {Bouma}, {Cartwright}, {Darveau-Bernier}, {Grankin}, {Ikoma}, {Kagetani}, {Kawauchi}, {Kodama}, {Kotani}, {Latham}, {Menou}, {Ricker}, {Seager}, {Tamura}, {Vanderspek}, \& {Watanabe}}]{Cadieux2022}
{Cadieux}, C., {Doyon}, R., {Plotnykov}, M., {et~al.} 2022, \href{http://dx.doi.org/10.3847/1538-3881/ac7cea}{\color{magenta}\aj}, \href{https://ui.adsabs.harvard.edu/abs/2022AJ....164...96C}{164, 96}

\bibitem[{{Cadieux} {et~al.}(2024){Cadieux}, {Plotnykov}, {Doyon}, {Valencia}, {Jahandar}, {Dang}, {Turbet}, {Fauchez}, {Cloutier}, {Cherubim}, {Artigau}, {Cook}, {Edwards}, {Hallatt}, {Charnay}, {Bouchy}, {Allart}, {Mignon}, {Baron}, {Barros}, {Benneke}, {Canto Martins}, {Cowan}, {De Medeiros}, {Delfosse}, {Delgado-Mena}, {Dumusque}, {Ehrenreich}, {Frensch}, {Gonz{\'a}lez Hern{\'a}ndez}, {Hara}, {Lafreni{\`e}re}, {Lo Curto}, {Malo}, {Melo}, {Mounzer}, {Passeger}, {Pepe}, {Poulin-Girard}, {Santos}, {Sosnowska}, {Su{\'a}rez Mascare{\~n}o}, {Thibault}, {Vaulato}, {Wade}, \& {Wildi}}]{Cadieux2024}
{Cadieux}, C., {Plotnykov}, M., {Doyon}, R., {et~al.} 2024, \href{http://dx.doi.org/10.3847/2041-8213/ad1691}{\color{magenta}\apjl}, \href{https://ui.adsabs.harvard.edu/abs/2024ApJ...960L...3C}{960, L3}

\bibitem[{{Chatterjee} \& {Pierrehumbert}(2024)}]{Chatterjee2024}
{Chatterjee}, R.~D. \& {Pierrehumbert}, R.~T. 2024, \href{https://ui.adsabs.harvard.edu/abs/2024arXiv241205188C}{\href{http://dx.doi.org/10.48550/arXiv.2412.05188}{\color{magenta}arXiv e-prints}, arXiv:2412.05188}

\bibitem[{{Choi} {et~al.}(2016){Choi}, {Dotter}, {Conroy}, {Cantiello}, {Paxton}, \& {Johnson}}]{Choi2016}
{Choi}, J., {Dotter}, A., {Conroy}, C., {et~al.} 2016, \href{http://dx.doi.org/10.3847/0004-637X/823/2/102}{\color{magenta}\apj}, \href{https://ui.adsabs.harvard.edu/abs/2016ApJ...823..102C}{823, 102}

\bibitem[{{Cloutier} {et~al.}(2019){Cloutier}, {Astudillo-Defru}, {Bonfils}, {Jenkins}, {Berdi{\~n}as}, {Ricker}, {Vanderspek}, {Latham}, {Seager}, {Winn}, {Jenkins}, {Almenara}, {Bouchy}, {Delfosse}, {D{\'\i}az}, {D{\'\i}az}, {Doyon}, {Figueira}, {Forveille}, {Kurtovic}, {Lovis}, {Mayor}, {Menou}, {Morgan}, {Morris}, {Muirhead}, {Murgas}, {Pepe}, {Santos}, {S{\'e}gransan}, {Smith}, {Tenenbaum}, {Torres}, {Udry}, {Vezie}, \& {Villasenor}}]{Cloutier2019}
{Cloutier}, R., {Astudillo-Defru}, N., {Bonfils}, X., {et~al.} 2019, \href{http://dx.doi.org/10.1051/0004-6361/201935957}{\color{magenta}\aap}, \href{https://ui.adsabs.harvard.edu/abs/2019A&A...629A.111C}{629, A111}

\bibitem[{{Crossfield} {et~al.}(2022){Crossfield}, {Malik}, {Hill}, {Kane}, {Foley}, {Polanski}, {Coria}, {Brande}, {Zhang}, {Wienke}, {Kreidberg}, {Cowan}, {Dragomir}, {Gorjian}, {Mikal-Evans}, {Benneke}, {Christiansen}, {Deming}, \& {Morales}}]{Crossfield2022}
{Crossfield}, I. J.~M., {Malik}, M., {Hill}, M.~L., {et~al.} 2022, \href{http://dx.doi.org/10.3847/2041-8213/ac886b}{\color{magenta}\apjl}, \href{https://ui.adsabs.harvard.edu/abs/2022ApJ...937L..17C}{937, L17}

\bibitem[{{Demangeon} {et~al.}(2021){Demangeon}, {Zapatero Osorio}, {Alibert}, {Barros}, {Adibekyan}, {Tabernero}, {Antoniadis-Karnavas}, {Camacho}, {Su{\'a}rez Mascare{\~n}o}, {Oshagh}, {Micela}, {Sousa}, {Lovis}, {Pepe}, {Rebolo}, {Cristiani}, {Santos}, {Allart}, {Allende Prieto}, {Bossini}, {Bouchy}, {Cabral}, {Damasso}, {Di Marcantonio}, {D'Odorico}, {Ehrenreich}, {Faria}, {Figueira}, {G{\'e}nova Santos}, {Haldemann}, {Hara}, {Gonz{\'a}lez Hern{\'a}ndez}, {Lavie}, {Lillo-Box}, {Lo Curto}, {Martins}, {M{\'e}gevand}, {Mehner}, {Molaro}, {Nunes}, {Pall{\'e}}, {Pasquini}, {Poretti}, {Sozzetti}, \& {Udry}}]{Demangeon2021}
{Demangeon}, O.~D.~S., {Zapatero Osorio}, M.~R., {Alibert}, Y., {et~al.} 2021, \href{http://dx.doi.org/10.1051/0004-6361/202140728}{\color{magenta}\aap}, \href{https://ui.adsabs.harvard.edu/abs/2021A&A...653A..41D}{653, A41}

\bibitem[{{Dotter}(2016)}]{Dotter2016}
{Dotter}, A. 2016, \href{http://dx.doi.org/10.3847/0067-0049/222/1/8}{\color{magenta}\apjs}, \href{https://ui.adsabs.harvard.edu/abs/2016ApJS..222....8D}{222, 8}

\bibitem[{{Douglas} {et~al.}(2019){Douglas}, {Curtis}, {Ag{\"u}eros}, {Cargile}, {Brewer}, {Meibom}, \& {Jansen}}]{Douglas2019}
{Douglas}, S.~T., {Curtis}, J.~L., {Ag{\"u}eros}, M.~A., {et~al.} 2019, \href{http://dx.doi.org/10.3847/1538-4357/ab2468}{\color{magenta}\apj}, \href{https://ui.adsabs.harvard.edu/abs/2019ApJ...879..100D}{879, 100}

\bibitem[{{Dungee} {et~al.}(2022){Dungee}, {van Saders}, {Gaidos}, {Chun}, {Garc{\'\i}a}, {Magnier}, {Mathur}, \& {Santos}}]{Dungee2022}
{Dungee}, R., {van Saders}, J., {Gaidos}, E., {et~al.} 2022, \href{http://dx.doi.org/10.3847/1538-4357/ac90be}{\color{magenta}\apj}, \href{https://ui.adsabs.harvard.edu/abs/2022ApJ...938..118D}{938, 118}

\bibitem[{{Engle}(2024)}]{Engle2024}
{Engle}, S.~G. 2024, \href{http://dx.doi.org/10.3847/1538-4357/ad0840}{\color{magenta}\apj}, \href{https://ui.adsabs.harvard.edu/abs/2024ApJ...960...62E}{960, 62}

\bibitem[{{Engle} \& {Guinan}(2023)}]{Engle2023}
{Engle}, S.~G. \& {Guinan}, E.~F. 2023, \href{http://dx.doi.org/10.3847/2041-8213/acf472}{\color{magenta}\apjl}, \href{https://ui.adsabs.harvard.edu/abs/2023ApJ...954L..50E}{954, L50}

\bibitem[{{Fantin} {et~al.}(2019){Fantin}, {C{\^o}t{\'e}}, {McConnachie}, {Bergeron}, {Cuillandre}, {Gwyn}, {Ibata}, {Thomas}, {Carlberg}, {Fabbro}, {Haywood}, {Lan{\c{c}}on}, {Lewis}, {Malhan}, {Martin}, {Navarro}, {Scott}, \& {Starkenburg}}]{Fantin2019}
{Fantin}, N.~J., {C{\^o}t{\'e}}, P., {McConnachie}, A.~W., {et~al.} 2019, \href{http://dx.doi.org/10.3847/1538-4357/ab5521}{\color{magenta}\apj}, \href{https://ui.adsabs.harvard.edu/abs/2019ApJ...887..148F}{887, 148}

\bibitem[{{France} {et~al.}(2018){France}, {Arulanantham}, {Fossati}, {Lanza}, {Loyd}, {Redfield}, \& {Schneider}}]{France2018}
{France}, K., {Arulanantham}, N., {Fossati}, L., {et~al.} 2018, \href{http://dx.doi.org/10.3847/1538-4365/aae1a3}{\color{magenta}\apjs}, \href{https://ui.adsabs.harvard.edu/abs/2018ApJS..239...16F}{239, 16}

\bibitem[{{France} {et~al.}(2020){France}, {Duvvuri}, {Egan}, {Koskinen}, {Wilson}, {Youngblood}, {Froning}, {Brown}, {Alvarado-G{\'o}mez}, {Berta-Thompson}, {Drake}, {Garraffo}, {Kaltenegger}, {Kowalski}, {Linsky}, {Loyd}, {Mauas}, {Miguel}, {Pineda}, {Rugheimer}, {Schneider}, {Tian}, \& {Vieytes}}]{France2020}
{France}, K., {Duvvuri}, G., {Egan}, H., {et~al.} 2020, \href{http://dx.doi.org/10.3847/1538-3881/abb465}{\color{magenta}\aj}, \href{https://ui.adsabs.harvard.edu/abs/2020AJ....160..237F}{160, 237}

\bibitem[{{Fuhrmann} {et~al.}(2012){Fuhrmann}, {Chini}, {Hoffmeister}, \& {Bernkopf}}]{Fuhrmann2012}
{Fuhrmann}, K., {Chini}, R., {Hoffmeister}, V.~H., \& {Bernkopf}, J. 2012, \href{http://dx.doi.org/10.1111/j.1365-2966.2011.20127.x}{\color{magenta}\mnras}, \href{https://ui.adsabs.harvard.edu/abs/2012MNRAS.420.1423F}{420, 1423}

\bibitem[{{Gaia Collaboration} {et~al.}(2023){Gaia Collaboration}, {Vallenari}, {Brown}, {Prusti}, {de Bruijne}, {Arenou}, {Babusiaux}, {Biermann}, {Creevey}, {Ducourant}, {Evans}, {Eyer}, {Guerra}, {Hutton}, {Jordi}, {Klioner}, {Lammers}, {Lindegren}, {Luri}, {Mignard}, {Panem}, {Pourbaix}, {Randich}, {Sartoretti}, {Soubiran}, {Tanga}, {Walton}, {Bailer-Jones}, {Bastian}, {Drimmel}, {Jansen}, {Katz}, {Lattanzi}, {van Leeuwen}, {Bakker}, {Cacciari}, {Casta{\~n}eda}, {De Angeli}, {Fabricius}, {Fouesneau}, {Fr{\'e}mat}, {Galluccio}, {Guerrier}, {Heiter}, {Masana}, {Messineo}, {Mowlavi}, {Nicolas}, {Nienartowicz}, {Pailler}, {Panuzzo}, {Riclet}, {Roux}, {Seabroke}, {Sordo}, {Th{\'e}venin}, {Gracia-Abril}, {Portell}, {Teyssier}, {Altmann}, {Andrae}, {Audard}, {Bellas-Velidis}, {Benson}, {Berthier}, {Blomme}, {Burgess}, {Busonero}, {Busso}, {C{\'a}novas}, {Carry}, {Cellino}, {Cheek}, {Clementini}, {Damerdji}, {Davidson}, {de Teodoro}, {Nu{\~n}ez Campos}, {Delchambre}, {Dell'Oro}, {Esquej},
  {Fern{\'a}ndez-Hern{\'a}ndez}, {Fraile}, {Garabato}, {Garc{\'\i}a-Lario}, {Gosset}, {Haigron}, {Halbwachs}, {Hambly}, {Harrison}, {Hern{\'a}ndez}, {Hestroffer}, {Hodgkin}, {Holl}, {Jan{\ss}en}, {Jevardat de Fombelle}, {Jordan}, {Krone-Martins}, {Lanzafame}, {L{\"o}ffler}, {Marchal}, {Marrese}, {Moitinho}, {Muinonen}, {Osborne}, {Pancino}, {Pauwels}, {Recio-Blanco}, {Reyl{\'e}}, {Riello}, {Rimoldini}, {Roegiers}, {Rybizki}, {Sarro}, {Siopis}, {Smith}, {Sozzetti}, {Utrilla}, {van Leeuwen}, {Abbas}, {{\'A}brah{\'a}m}, {Abreu Aramburu}, {Aerts}, {Aguado}, {Ajaj}, {Aldea-Montero}, {Altavilla}, {{\'A}lvarez}, {Alves}, {Anders}, {Anderson}, {Anglada Varela}, {Antoja}, {Baines}, {Baker}, {Balaguer-N{\'u}{\~n}ez}, {Balbinot}, {Balog}, {Barache}, {Barbato}, {Barros}, {Barstow}, {Bartolom{\'e}}, {Bassilana}, {Bauchet}, {Becciani}, {Bellazzini}, {Berihuete}, {Bernet}, {Bertone}, {Bianchi}, {Binnenfeld}, {Blanco-Cuaresma}, {Blazere}, {Boch}, {Bombrun}, {Bossini}, {Bouquillon}, {Bragaglia}, {Bramante}, {Breedt},
  {Bressan}, {Brouillet}, {Brugaletta}, {Bucciarelli}, {Burlacu}, {Butkevich}, {Buzzi}, {Caffau}, {Cancelliere}, {Cantat-Gaudin}, {Carballo}, {Carlucci}, {Carnerero}, {Carrasco}, {Casamiquela}, {Castellani}, {Castro-Ginard}, {Chaoul}, {Charlot}, {Chemin}, {Chiaramida}, {Chiavassa}, {Chornay}, {Comoretto}, {Contursi}, {Cooper}, {Cornez}, {Cowell}, {Crifo}, {Cropper}, {Crosta}, {Crowley}, {Dafonte}, {Dapergolas}, {David}, {David}, {de Laverny}, {De Luise}, {De March}, {De Ridder}, {de Souza}, {de Torres}, {del Peloso}, {del Pozo}, {Delbo}, {Delgado}, {Delisle}, {Demouchy}, {Dharmawardena}, {Di Matteo}, {Diakite}, {Diener}, {Distefano}, {Dolding}, {Edvardsson}, {Enke}, {Fabre}, {Fabrizio}, {Faigler}, {Fedorets}, {Fernique}, {Fienga}, {Figueras}, {Fournier}, {Fouron}, {Fragkoudi}, {Gai}, {Garcia-Gutierrez}, {Garcia-Reinaldos}, {Garc{\'\i}a-Torres}, {Garofalo}, {Gavel}, {Gavras}, {Gerlach}, {Geyer}, {Giacobbe}, {Gilmore}, {Girona}, {Giuffrida}, {Gomel}, {Gomez}, {Gonz{\'a}lez-N{\'u}{\~n}ez},
  {Gonz{\'a}lez-Santamar{\'\i}a}, {Gonz{\'a}lez-Vidal}, {Granvik}, {Guillout}, {Guiraud}, {Guti{\'e}rrez-S{\'a}nchez}, {Guy}, {Hatzidimitriou}, {Hauser}, {Haywood}, {Helmer}, {Helmi}, {Sarmiento}, {Hidalgo}, {Hilger}, {H{\l}adczuk}, {Hobbs}, {Holland}, {Huckle}, {Jardine}, {Jasniewicz}, {Jean-Antoine Piccolo}, {Jim{\'e}nez-Arranz}, {Jorissen}, {Juaristi Campillo}, {Julbe}, {Karbevska}, {Kervella}, {Khanna}, {Kontizas}, {Kordopatis}, {Korn}, {K{\'o}sp{\'a}l}, {Kostrzewa-Rutkowska}, {Kruszy{\'n}ska}, {Kun}, {Laizeau}, {Lambert}, {Lanza}, {Lasne}, {Le Campion}, {Lebreton}, {Lebzelter}, {Leccia}, {Leclerc}, {Lecoeur-Taibi}, {Liao}, {Licata}, {Lindstr{\o}m}, {Lister}, {Livanou}, {Lobel}, {Lorca}, {Loup}, {Madrero Pardo}, {Magdaleno Romeo}, {Managau}, {Mann}, {Manteiga}, {Marchant}, {Marconi}, {Marcos}, {Marcos Santos}, {Mar{\'\i}n Pina}, {Marinoni}, {Marocco}, {Marshall}, {Martin Polo}, {Mart{\'\i}n-Fleitas}, {Marton}, {Mary}, {Masip}, {Massari}, {Mastrobuono-Battisti}, {Mazeh}, {McMillan}, {Messina}, {Michalik},
  {Millar}, {Mints}, {Molina}, {Molinaro}, {Moln{\'a}r}, {Monari}, {Mongui{\'o}}, {Montegriffo}, {Montero}, {Mor}, {Mora}, {Morbidelli}, {Morel}, {Morris}, {Muraveva}, {Murphy}, {Musella}, {Nagy}, {Noval}, {Oca{\~n}a}, {Ogden}, {Ordenovic}, {Osinde}, {Pagani}, {Pagano}, {Palaversa}, {Palicio}, {Pallas-Quintela}, {Panahi}, {Payne-Wardenaar}, {Pe{\~n}alosa Esteller}, {Penttil{\"a}}, {Pichon}, {Piersimoni}, {Pineau}, {Plachy}, {Plum}, {Poggio}, {Pr{\v{s}}a}, {Pulone}, {Racero}, {Ragaini}, {Rainer}, {Raiteri}, {Rambaux}, {Ramos}, {Ramos-Lerate}, {Re Fiorentin}, {Regibo}, {Richards}, {Rios Diaz}, {Ripepi}, {Riva}, {Rix}, {Rixon}, {Robichon}, {Robin}, {Robin}, {Roelens}, {Rogues}, {Rohrbasser}, {Romero-G{\'o}mez}, {Rowell}, {Royer}, {Ruz Mieres}, {Rybicki}, {Sadowski}, {S{\'a}ez N{\'u}{\~n}ez}, {Sagrist{\`a} Sell{\'e}s}, {Sahlmann}, {Salguero}, {Samaras}, {Sanchez Gimenez}, {Sanna}, {Santove{\~n}a}, {Sarasso}, {Schultheis}, {Sciacca}, {Segol}, {Segovia}, {S{\'e}gransan}, {Semeux}, {Shahaf}, {Siddiqui}, {Siebert},
  {Siltala}, {Silvelo}, {Slezak}, {Slezak}, {Smart}, {Snaith}, {Solano}, {Solitro}, {Souami}, {Souchay}, {Spagna}, {Spina}, {Spoto}, {Steele}, {Steidelm{\"u}ller}, {Stephenson}, {S{\"u}veges}, {Surdej}, {Szabados}, {Szegedi-Elek}, {Taris}, {Taylor}, {Teixeira}, {Tolomei}, {Tonello}, {Torra}, {Torra}, {Torralba Elipe}, {Trabucchi}, {Tsounis}, {Turon}, {Ulla}, {Unger}, {Vaillant}, {van Dillen}, {van Reeven}, {Vanel}, {Vecchiato}, {Viala}, {Vicente}, {Voutsinas}, {Weiler}, {Wevers}, {Wyrzykowski}, {Yoldas}, {Yvard}, {Zhao}, {Zorec}, {Zucker}, \& {Zwitter}}]{Gaia2023}
{Gaia Collaboration}, {Vallenari}, A., {Brown}, A.~G.~A., {et~al.} 2023, \href{http://dx.doi.org/10.1051/0004-6361/202243940}{\color{magenta}\aap}, \href{https://ui.adsabs.harvard.edu/abs/2023A&A...674A...1G}{674, A1}

\bibitem[{{Ghachoui} {et~al.}(2024){Ghachoui}, {Rackham}, {D{\'e}vora-Pajares}, {Chouqar}, {Timmermans}, {Kaltenegger}, {Sebastian}, {Pozuelos}, {Eastman}, {Burgasser}, {Murgas}, {Stassun}, {Gillon}, {Benkhaldoun}, {Palle}, {Delrez}, {Jenkins}, {Barkaoui}, {Narita}, {de Leon}, {Mori}, {Shporer}, {Rowden}, {Kostov}, {F{\H{u}}r{\'e}sz}, {Collins}, {Schwarz}, {Charbonneau}, {Guerrero}, {Ricker}, {Jehin}, {Fukui}, {Kawai}, {Hayashi}, {Esparza-Borges}, {Parviainen}, {Clark}, {Ciardi}, {Polanski}, {Schleider}, {Gilbert}, {Crossfield}, {Barclay}, {Dressing}, {Karpoor}, {Softich}, {Gerasimov}, \& {Davoudi}}]{Ghachoui2024}
{Ghachoui}, M., {Rackham}, B.~V., {D{\'e}vora-Pajares}, M., {et~al.} 2024, \href{http://dx.doi.org/10.1051/0004-6361/202451120}{\color{magenta}\aap}, \href{https://ui.adsabs.harvard.edu/abs/2024A&A...690A.263G}{690, A263}

\bibitem[{{Gillon} {et~al.}(2024){Gillon}, {Pedersen}, {Rackham}, {Dransfield}, {Ducrot}, {Barkaoui}, {Burdanov}, {Schroffenegger}, {G{\'o}mez Maqueo Chew}, {Lederer}, {Alonso}, {Burgasser}, {Howell}, {Narita}, {de Wit}, {Demory}, {Queloz}, {Triaud}, {Delrez}, {Jehin}, {Hooton}, {Garcia}, {Jano Mu{\~n}oz}, {Murray}, {Pozuelos}, {Sebastian}, {Timmermans}, {Thompson}, {Z{\'u}{\~n}iga-Fern{\'a}ndez}, {Aceituno}, {Aganze}, {Amado}, {Baycroft}, {Benkhaldoun}, {Berardo}, {Bolmont}, {Clark}, {Davis}, {Davoudi}, {de Beurs}, {de Leon}, {Ikoma}, {Ikuta}, {Isogai}, {Fukuda}, {Fukui}, {Gerasimov}, {Ghachoui}, {G{\"u}nther}, {Hasler}, {Hayashi}, {Heng}, {Hu}, {Kagetani}, {Kawai}, {Kawauchi}, {Kitzmann}, {Koll}, {Lendl}, {Livingston}, {Lyu}, {Meier Vald{\'e}s}, {Mori}, {McCormac}, {Murgas}, {Niraula}, {Pall{\'e}}, {Plauchu-Frayn}, {Rebolo}, {Sabin}, {Schackey}, {Schanche}, {Selsis}, {Sota}, {Stalport}, {Standing}, {Stassun}, {Tamura}, {Terada}, {Theissen}, {Turbet}, {Van Grootel}, {Varas}, {Watanabe}, \& {Zong
  Lang}}]{Gillon2024}
{Gillon}, M., {Pedersen}, P.~P., {Rackham}, B.~V., {et~al.} 2024, \href{http://dx.doi.org/10.1038/s41550-024-02271-2}{\color{magenta}Nature Astronomy}, \href{https://ui.adsabs.harvard.edu/abs/2024NatAs...8..865G}{8, 865}

\bibitem[{{Greene} {et~al.}(2023){Greene}, {Bell}, {Ducrot}, {Dyrek}, {Lagage}, \& {Fortney}}]{Greene2023}
{Greene}, T.~P., {Bell}, T.~J., {Ducrot}, E., {et~al.} 2023, \href{http://dx.doi.org/10.1038/s41586-023-05951-7}{\color{magenta}\nat}, \href{https://ui.adsabs.harvard.edu/abs/2023Natur.618...39G}{618, 39}

\bibitem[{Harris {et~al.}(2020)Harris, Millman, van~der Walt, Gommers, Virtanen, Cournapeau, Wieser, Taylor, Berg, Smith, Kern, Picus, Hoyer, van Kerkwijk, Brett, Haldane, del R{\'{i}}o, Wiebe, Peterson, G{\'{e}}rard-Marchant, Sheppard, Reddy, Weckesser, Abbasi, Gohlke, \& Oliphant}]{Harris2020}
Harris, C.~R., Millman, K.~J., van~der Walt, S.~J., {et~al.} 2020, \href{http://dx.doi.org/10.1038/s41586-020-2649-2}{\color{magenta}Nature}, \href{https://ui.adsabs.harvard.edu/abs/2020Natur.585..357H}{585, 357}

\bibitem[{{Hirano} {et~al.}(2023){Hirano}, {Dai}, {Livingston}, {Grziwa}, {Lam}, {Kasagi}, {Narita}, {Ishikawa}, {Miyakawa}, {Serrano}, {Matsumoto}, {Kokubo}, {Kimura}, {Ikoma}, {Winn}, {Wisniewski}, {Harakawa}, {Teng}, {Cochran}, {Fukui}, {Gandolfi}, {Guenther}, {Hori}, {Ikuta}, {Kawauchi}, {Knudstrup}, {Korth}, {Kotani}, {Krishnamurthy}, {Kudo}, {Kurokawa}, {Kuzuhara}, {Luque}, {Mori}, {Nishikawa}, {Omiya}, {Orell-Miquel}, {Palle}, {Persson}, {Redfield}, {Serabyn}, {Smith}, {Takahashi}, {Takarada}, {Ueda}, {Van Eylen}, {Vievard}, {Tamura}, \& {Sato}}]{Hirano2023}
{Hirano}, T., {Dai}, F., {Livingston}, J.~H., {et~al.} 2023, \href{http://dx.doi.org/10.3847/1538-3881/acb7e1}{\color{magenta}\aj}, \href{https://ui.adsabs.harvard.edu/abs/2023AJ....165..131H}{165, 131}

\bibitem[{Hunter(2007)}]{Hunter2007}
Hunter, J.~D. 2007, \href{http://dx.doi.org/10.1109/MCSE.2007.55}{\color{magenta}CSE}, \href{https://ui.adsabs.harvard.edu/abs/2007CSE.....9...90H}{9, 90}

\bibitem[{{Irwin} {et~al.}(2015){Irwin}, {Berta-Thompson}, {Charbonneau}, {Dittmann}, {Falco}, {Newton}, \& {Nutzman}}]{Irwin2015}
{Irwin}, J.~M., {Berta-Thompson}, Z.~K., {Charbonneau}, D., {et~al.} 2015, in 18th Cambridge Workshop on Cool Stars, Stellar Systems, and the Sun, ed. G.~van Belle \& H.~C. Harris (Flagstaff, AZ: Lowell Observatory), \href{https://ui.adsabs.harvard.edu/abs/2015csss...18..767I}{767}

\bibitem[{{Jardine} \& {Unruh}(1999)}]{Jardine1999}
{Jardine}, M. \& {Unruh}, Y.~C. 1999, \aap, \href{https://ui.adsabs.harvard.edu/abs/1999A&A...346..883J}{346, 883}

\bibitem[{{Johnstone} {et~al.}(2021){Johnstone}, {Bartel}, \& {G{\"u}del}}]{Johnstone2021}
{Johnstone}, C.~P., {Bartel}, M., \& {G{\"u}del}, M. 2021, \href{http://dx.doi.org/10.1051/0004-6361/202038407}{\color{magenta}\aap}, \href{https://ui.adsabs.harvard.edu/abs/2021A&A...649A..96J}{649, A96}

\bibitem[{{King} \& {Wheatley}(2021)}]{King2021}
{King}, G.~W. \& {Wheatley}, P.~J. 2021, \href{http://dx.doi.org/10.1093/mnrasl/slaa186}{\color{magenta}\mnras}, \href{https://ui.adsabs.harvard.edu/abs/2021MNRAS.501L..28K}{501, L28}

\bibitem[{{Kreidberg} {et~al.}(2019){Kreidberg}, {Koll}, {Morley}, {Hu}, {Schaefer}, {Deming}, {Stevenson}, {Dittmann}, {Vanderburg}, {Berardo}, {Guo}, {Stassun}, {Crossfield}, {Charbonneau}, {Latham}, {Loeb}, {Ricker}, {Seager}, \& {Vanderspek}}]{Kreidberg2019}
{Kreidberg}, L., {Koll}, D. D.~B., {Morley}, C., {et~al.} 2019, \href{http://dx.doi.org/10.1038/s41586-019-1497-4}{\color{magenta}\nat}, \href{https://ui.adsabs.harvard.edu/abs/2019Natur.573...87K}{573, 87}

\bibitem[{{Lammer} {et~al.}(2009){Lammer}, {Odert}, {Leitzinger}, {Khodachenko}, {Panchenko}, {Kulikov}, {Zhang}, {Lichtenegger}, {Erkaev}, {Wuchterl}, {Micela}, {Penz}, {Biernat}, {Weingrill}, {Steller}, {Ottacher}, {Hasiba}, \& {Hanslmeier}}]{Lammer2009}
{Lammer}, H., {Odert}, P., {Leitzinger}, M., {et~al.} 2009, \href{http://dx.doi.org/10.1051/0004-6361/200911922}{\color{magenta}\aap}, \href{https://ui.adsabs.harvard.edu/abs/2009A&A...506..399L}{506, 399}

\bibitem[{{Lehtinen} {et~al.}(2020){Lehtinen}, {Spada}, {K{\"a}pyl{\"a}}, {Olspert}, \& {K{\"a}pyl{\"a}}}]{Lehtinen2020}
{Lehtinen}, J.~J., {Spada}, F., {K{\"a}pyl{\"a}}, M.~J., {Olspert}, N., \& {K{\"a}pyl{\"a}}, P.~J. 2020, \href{http://dx.doi.org/10.1038/s41550-020-1039-x}{\color{magenta}Nature Astronomy}, \href{https://ui.adsabs.harvard.edu/abs/2020NatAs...4..658L}{4, 658}

\bibitem[{{Linsky} {et~al.}(2020){Linsky}, {Wood}, {Youngblood}, {Brown}, {Froning}, {France}, {Buccino}, {Cranmer}, {Mauas}, {Miguel}, {Pineda}, {Rugheimer}, {Vieytes}, {Wheatley}, \& {Wilson}}]{Linsky2020}
{Linsky}, J.~L., {Wood}, B.~E., {Youngblood}, A., {et~al.} 2020, \href{http://dx.doi.org/10.3847/1538-4357/abb36f}{\color{magenta}\apj}, \href{https://ui.adsabs.harvard.edu/abs/2020ApJ...902....3L}{902, 3}

\bibitem[{{Luger} \& {Barnes}(2015)}]{Luger2015}
{Luger}, R. \& {Barnes}, R. 2015, \href{http://dx.doi.org/10.1089/ast.2014.1231}{\color{magenta}Astrobiology}, \href{https://ui.adsabs.harvard.edu/abs/2015AsBio..15..119L}{15, 119}

\bibitem[{{Luque} {et~al.}(2024){Luque}, {Park Coy}, {Xue}, {Feinstein}, {Ahrer}, {Changeat}, {Zhang}, {Moran}, {Bean}, {Kite}, {Weiner Mansfield}, \& {Pall{\'e}}}]{Luque2024}
{Luque}, R., {Park Coy}, B., {Xue}, Q., {et~al.} 2024, \href{https://ui.adsabs.harvard.edu/abs/2024arXiv241203411L}{\href{http://dx.doi.org/10.48550/arXiv.2412.03411}{\color{magenta}arXiv e-prints}, arXiv:2412.03411}

\bibitem[{{Mann} {et~al.}(2015){Mann}, {Feiden}, {Gaidos}, {Boyajian}, \& {von Braun}}]{Mann2015}
{Mann}, A.~W., {Feiden}, G.~A., {Gaidos}, E., {Boyajian}, T., \& {von Braun}, K. 2015, \href{http://dx.doi.org/10.1088/0004-637X/804/1/64}{\color{magenta}\apj}, \href{https://ui.adsabs.harvard.edu/abs/2015ApJ...804...64M}{804, 64}

\bibitem[{{Medina} {et~al.}(2020){Medina}, {Winters}, {Irwin}, \& {Charbonneau}}]{Medina2020}
{Medina}, A.~A., {Winters}, J.~G., {Irwin}, J.~M., \& {Charbonneau}, D. 2020, \href{http://dx.doi.org/10.3847/1538-4357/abc686}{\color{magenta}ApJ}, \href{https://ui.adsabs.harvard.edu/abs/2020ApJ...905..107M}{905, 107}

\bibitem[{{Medina} {et~al.}(2022){Medina}, {Winters}, {Irwin}, \& {Charbonneau}}]{Medina2022_flare}
{Medina}, A.~A., {Winters}, J.~G., {Irwin}, J.~M., \& {Charbonneau}, D. 2022, \href{http://dx.doi.org/10.3847/1538-4357/ac77f9}{\color{magenta}\apj}, \href{https://ui.adsabs.harvard.edu/abs/2022ApJ...935..104M}{935, 104}

\bibitem[{{Meier Valdes} {et~al.}(2025){Meier Valdes}, {Demory}, {Diamond-Lowe}, {Mendonça}, {August}, {Fortune}, {Allen}, {Kitzmann}, {Gressier}, {Hooton}, {Jones}, {Buchhave}, {Espinoza}, {Fisher}, {Gibson}, {Heng}, {Hoeijmakers}, {Prinoth}, {Rathcke}, \& {Eastman}}]{MeierValdes2025}
{Meier Valdes}, E.~A., {Demory}, B.~O., {Diamond-Lowe}, H., {et~al.} 2025, \href{https://ui.adsabs.harvard.edu/abs/2025arXiv250319772M}{arXiv e-prints, arXiv:2503.19772}

\bibitem[{{Ment} {et~al.}(2019){Ment}, {Dittmann}, {Astudillo-Defru}, {Charbonneau}, {Irwin}, {Bonfils}, {Murgas}, {Almenara}, {Forveille}, {Agol}, {Ballard}, {Berta-Thompson}, {Bouchy}, {Cloutier}, {Delfosse}, {Doyon}, {Dressing}, {Esquerdo}, {Haywood}, {Kipping}, {Latham}, {Lovis}, {Newton}, {Pepe}, {Rodriguez}, {Santos}, {Tan}, {Udry}, {Winters}, \& {W{\"u}nsche}}]{Ment2019}
{Ment}, K., {Dittmann}, J.~A., {Astudillo-Defru}, N., {et~al.} 2019, \href{http://dx.doi.org/10.3847/1538-3881/aaf1b1}{\color{magenta}\aj}, \href{https://ui.adsabs.harvard.edu/abs/2019AJ....157...32M}{157, 32}

\bibitem[{{Ment} {et~al.}(2021){Ment}, {Irwin}, {Charbonneau}, {Winters}, {Medina}, {Cloutier}, {D{\'\i}az}, {Jenkins}, {Ziegler}, {Law}, {Mann}, {Ricker}, {Vanderspek}, {Latham}, {Seager}, {Winn}, {Jenkins}, {Goeke}, {Levine}, {Rojas-Ayala}, {Rowden}, {Ting}, \& {Twicken}}]{Ment2021}
{Ment}, K., {Irwin}, J., {Charbonneau}, D., {et~al.} 2021, \href{http://dx.doi.org/10.3847/1538-3881/abbd91}{\color{magenta}\aj}, \href{https://ui.adsabs.harvard.edu/abs/2021AJ....161...23M}{161, 23}

\bibitem[{{Mohanty} {et~al.}(2002){Mohanty}, {Basri}, {Shu}, {Allard}, \& {Chabrier}}]{Mohanty2002}
{Mohanty}, S., {Basri}, G., {Shu}, F., {Allard}, F., \& {Chabrier}, G. 2002, \href{http://dx.doi.org/10.1086/339911}{\color{magenta}\apj}, \href{https://ui.adsabs.harvard.edu/abs/2002ApJ...571..469M}{571, 469}

\bibitem[{{Moore} \& {Cowan}(2020)}]{Moore2020}
{Moore}, K. \& {Cowan}, N.~B. 2020, \href{http://dx.doi.org/10.1093/mnras/staa1796}{\color{magenta}\mnras}, \href{https://ui.adsabs.harvard.edu/abs/2020MNRAS.496.3786M}{496, 3786}

\bibitem[{{Nakayama} {et~al.}(2022){Nakayama}, {Ikoma}, \& {Terada}}]{Nakayama2022}
{Nakayama}, A., {Ikoma}, M., \& {Terada}, N. 2022, \href{http://dx.doi.org/10.3847/1538-4357/ac86ca}{\color{magenta}\apj}, \href{https://ui.adsabs.harvard.edu/abs/2022ApJ...937...72N}{937, 72}

\bibitem[{{Newton} {et~al.}(2017){Newton}, {Irwin}, {Charbonneau}, {Berlind}, {Calkins}, \& {Mink}}]{Newton2017}
{Newton}, E.~R., {Irwin}, J., {Charbonneau}, D., {et~al.} 2017, \href{http://dx.doi.org/10.3847/1538-4357/834/1/85}{\color{magenta}\apj}, \href{https://ui.adsabs.harvard.edu/abs/2017ApJ...834...85N}{834, 85}

\bibitem[{{Newton} {et~al.}(2016){Newton}, {Irwin}, {Charbonneau}, {Berta-Thompson}, {Dittmann}, \& {West}}]{Newton2016}
{Newton}, E.~R., {Irwin}, J., {Charbonneau}, D., {et~al.} 2016, \href{http://dx.doi.org/10.3847/0004-637X/821/2/93}{\color{magenta}\apj}, \href{https://ui.adsabs.harvard.edu/abs/2016ApJ...821...93N}{821, 93}

\bibitem[{{Newton} {et~al.}(2018){Newton}, {Mondrik}, {Irwin}, {Winters}, \& {Charbonneau}}]{Newton2018}
{Newton}, E.~R., {Mondrik}, N., {Irwin}, J., {Winters}, J.~G., \& {Charbonneau}, D. 2018, \href{http://dx.doi.org/10.3847/1538-3881/aad73b}{\color{magenta}\aj}, \href{https://ui.adsabs.harvard.edu/abs/2018AJ....156..217N}{156, 217}

\bibitem[{{Noyes} {et~al.}(1984){Noyes}, {Hartmann}, {Baliunas}, {Duncan}, \& {Vaughan}}]{Noyes1984}
{Noyes}, R.~W., {Hartmann}, L.~W., {Baliunas}, S.~L., {Duncan}, D.~K., \& {Vaughan}, A.~H. 1984, \href{http://dx.doi.org/10.1086/161945}{\color{magenta}\apj}, \href{https://ui.adsabs.harvard.edu/abs/1984ApJ...279..763N}{279, 763}

\bibitem[{{Nutzman} \& {Charbonneau}(2008)}]{Nutzman2008}
{Nutzman}, P. \& {Charbonneau}, D. 2008, \href{http://dx.doi.org/10.1086/533420}{\color{magenta}\pasp}, \href{https://ui.adsabs.harvard.edu/abs/2008PASP..120..317N}{120, 317}

\bibitem[{{Park Coy} {et~al.}(2024){Park Coy}, {Ih}, {Kite}, {Koll}, {Tenthoff}, {Bean}, {Weiner Mansfield}, {Zhang}, {Xue}, {Kempton}, {Wolhfarth}, {Hu}, {Lyu}, \& {Wohler}}]{ParkCoy2024}
{Park Coy}, B., {Ih}, J., {Kite}, E.~S., {et~al.} 2024, \href{https://ui.adsabs.harvard.edu/abs/2024arXiv241206573P}{\href{http://dx.doi.org/10.48550/arXiv.2412.06573}{\color{magenta}arXiv e-prints}, arXiv:2412.06573}

\bibitem[{{Pass} {et~al.}(2022){Pass}, {Charbonneau}, {Irwin}, \& {Winters}}]{Pass2022}
{Pass}, E.~K., {Charbonneau}, D., {Irwin}, J.~M., \& {Winters}, J.~G. 2022, \href{http://dx.doi.org/10.3847/1538-4357/ac7da8}{\color{magenta}\apj}, \href{https://ui.adsabs.harvard.edu/abs/2022ApJ...936..109P}{936, 109}

\bibitem[{{Pass} {et~al.}(2024){Pass}, {Charbonneau}, {Latham}, {Berlind}, {Calkins}, {Esquerdo}, \& {Mink}}]{Pass2024}
{Pass}, E.~K., {Charbonneau}, D., {Latham}, D.~W., {et~al.} 2024, \href{http://dx.doi.org/10.3847/1538-4357/ad3631}{\color{magenta}\apj}, \href{https://ui.adsabs.harvard.edu/abs/2024ApJ...966..231P}{966, 231}

\bibitem[{{Pass} {et~al.}(2023{\natexlab{a}}){Pass}, {Winters}, {Charbonneau}, {Irwin}, {Latham}, {Berlind}, {Calkins}, {Esquerdo}, \& {Mink}}]{Pass2023_inactive}
{Pass}, E.~K., {Winters}, J.~G., {Charbonneau}, D., {et~al.} 2023{\natexlab{a}}, \href{http://dx.doi.org/10.3847/1538-3881/acd349}{\color{magenta}\aj}, \href{https://ui.adsabs.harvard.edu/abs/2023AJ....166...11P}{166, 11}

\bibitem[{{Pass} {et~al.}(2023{\natexlab{b}}){Pass}, {Winters}, {Charbonneau}, {Irwin}, \& {Medina}}]{Pass2023_active}
{Pass}, E.~K., {Winters}, J.~G., {Charbonneau}, D., {Irwin}, J.~M., \& {Medina}, A.~A. 2023{\natexlab{b}}, \href{http://dx.doi.org/10.3847/1538-3881/acd6a2}{\color{magenta}\aj}, \href{https://ui.adsabs.harvard.edu/abs/2023AJ....166...16P}{166, 16}

\bibitem[{Pecaut \& Mamajek(2013)}]{Pecaut2013}
Pecaut, M.~J. \& Mamajek, E.~E. 2013, \href{http://dx.doi.org/10.1088/0067-0049/208/1/9}{\color{magenta}ApJS}, \href{https://ui.adsabs.harvard.edu/abs/2013ApJS..208....9P}{208, 9}

\bibitem[{{Penz} \& {Micela}(2008)}]{Penz2008b}
{Penz}, T. \& {Micela}, G. 2008, \href{http://dx.doi.org/10.1051/0004-6361:20078873}{\color{magenta}\aap}, \href{https://ui.adsabs.harvard.edu/abs/2008A&A...479..579P}{479, 579}

\bibitem[{{Penz} {et~al.}(2008){Penz}, {Micela}, \& {Lammer}}]{Penz2008a}
{Penz}, T., {Micela}, G., \& {Lammer}, H. 2008, \href{http://dx.doi.org/10.1051/0004-6361:20078364}{\color{magenta}\aap}, \href{https://ui.adsabs.harvard.edu/abs/2008A&A...477..309P}{477, 309}

\bibitem[{{Pineda} {et~al.}(2021){Pineda}, {Youngblood}, \& {France}}]{Pineda2021}
{Pineda}, J.~S., {Youngblood}, A., \& {France}, K. 2021, \href{http://dx.doi.org/10.3847/1538-4357/abe8d7}{\color{magenta}\apj}, \href{https://ui.adsabs.harvard.edu/abs/2021ApJ...911..111P}{911, 111}

\bibitem[{{Reback} {et~al.}(2021){Reback}, {Jbrockmendel}, {McKinney}, {Van Den Bossche}, {Augspurger}, {Cloud}, {Hawkins}, {Gfyoung}, {Sinhrks}, {Roeschke}, {Klein}, {Petersen}, {Tratner}, {She}, {Ayd}, {Hoefler}, {Naveh}, {Garcia}, {Schendel}, {Hayden}, {Saxton}, {Gorelli}, {Shadrach}, {Jancauskas}, {McMaster}, {Li}, {Battiston}, {Seabold}, {Attack68}, \& {Dong}}]{Reback2021}
{Reback}, J., {Jbrockmendel}, {McKinney}, W., {et~al.} 2021, {pandas-dev/pandas: Pandas 1.3.0}, Zenodo, doi:10.5281/zenodo.3509134

\bibitem[{{Redfield} {et~al.}(2024){Redfield}, {Batalha}, {Benneke}, {Biller}, {Espinoza}, {France}, {Konopacky}, {Kreidberg}, {Rauscher}, \& {Sing}}]{Redfield2024}
{Redfield}, S., {Batalha}, N., {Benneke}, B., {et~al.} 2024, \href{https://ui.adsabs.harvard.edu/abs/2024arXiv240402932R}{\href{http://dx.doi.org/10.48550/arXiv.2404.02932}{\color{magenta}arXiv e-prints}, arXiv:2404.02932}

\bibitem[{{Reiners} \& {Basri}(2008)}]{Reiners2008}
{Reiners}, A. \& {Basri}, G. 2008, \href{http://dx.doi.org/10.1086/590073}{\color{magenta}\apj}, \href{https://ui.adsabs.harvard.edu/abs/2008ApJ...684.1390R}{684, 1390}

\bibitem[{{Ribas} {et~al.}(2016){Ribas}, {Bolmont}, {Selsis}, {Reiners}, {Leconte}, {Raymond}, {Engle}, {Guinan}, {Morin}, {Turbet}, {Forget}, \& {Anglada-Escud{\'e}}}]{Ribas2016}
{Ribas}, I., {Bolmont}, E., {Selsis}, F., {et~al.} 2016, \href{http://dx.doi.org/10.1051/0004-6361/201629576}{\color{magenta}\aap}, \href{https://ui.adsabs.harvard.edu/abs/2016A&A...596A.111R}{596, A111}

\bibitem[{{Ribas} {et~al.}(2005){Ribas}, {Guinan}, {G{\"u}del}, \& {Audard}}]{Ribas2005}
{Ribas}, I., {Guinan}, E.~F., {G{\"u}del}, M., \& {Audard}, M. 2005, \href{http://dx.doi.org/10.1086/427977}{\color{magenta}\apj}, \href{https://ui.adsabs.harvard.edu/abs/2005ApJ...622..680R}{622, 680}

\bibitem[{{Selsis} {et~al.}(2007){Selsis}, {Kasting}, {Levrard}, {Paillet}, {Ribas}, \& {Delfosse}}]{Selsis2007}
{Selsis}, F., {Kasting}, J.~F., {Levrard}, B., {et~al.} 2007, \href{http://dx.doi.org/10.1051/0004-6361:20078091}{\color{magenta}\aap}, \href{https://ui.adsabs.harvard.edu/abs/2007A&A...476.1373S}{476, 1373}

\bibitem[{{Skrutskie} {et~al.}(2006){Skrutskie}, {Cutri}, {Stiening}, {Weinberg}, {Schneider}, {Carpenter}, {Beichman}, {Capps}, {Chester}, {Elias}, {Huchra}, {Liebert}, {Lonsdale}, {Monet}, {Price}, {Seitzer}, {Jarrett}, {Kirkpatrick}, {Gizis}, {Howard}, {Evans}, {Fowler}, {Fullmer}, {Hurt}, {Light}, {Kopan}, {Marsh}, {McCallon}, {Tam}, {Van Dyk}, \& {Wheelock}}]{Skrutskie2006}
{Skrutskie}, M.~F., {Cutri}, R.~M., {Stiening}, R., {et~al.} 2006, \href{http://dx.doi.org/10.1086/498708}{\color{magenta}\aj}, \href{https://ui.adsabs.harvard.edu/abs/2006AJ....131.1163S}{131, 1163}

\bibitem[{{Skumanich}(1972)}]{Skumanich1972}
{Skumanich}, A. 1972, \href{http://dx.doi.org/10.1086/151310}{\color{magenta}\apj}, \href{https://ui.adsabs.harvard.edu/abs/1972ApJ...171..565S}{171, 565}

\bibitem[{{Soderblom}(2010)}]{Soderblom2010}
{Soderblom}, D.~R. 2010, \href{http://dx.doi.org/10.1146/annurev-astro-081309-130806}{\color{magenta}\araa}, \href{https://ui.adsabs.harvard.edu/abs/2010ARA&A..48..581S}{48, 581}

\bibitem[{{Terrien} {et~al.}(2022){Terrien}, {Keen}, {Oda}, {Parts(they/them)}, {Stef{\'a}nsson}, {Mahadevan}, {Robertson}, {Ninan}, {Beard}, {Bender}, {Cochran}, {Cunha}, {Diddams}, {Fredrick}, {Halverson}, {Hearty}, {Ickler}, {Kanodia}, {Libby-Roberts}, {Lubin}, {Metcalf}, {Olsen}, {Ramsey}, {Roy}, {Schwab}, {Smith}, \& {Turner}}]{Terrien2022}
{Terrien}, R.~C., {Keen}, A., {Oda}, K., {et~al.} 2022, \href{http://dx.doi.org/10.3847/2041-8213/ac4fc8}{\color{magenta}\apjl}, \href{https://ui.adsabs.harvard.edu/abs/2022ApJ...927L..11T}{927, L11}

\bibitem[{{Truemper}(1982)}]{Truemper1982}
{Truemper}, J. 1982, \href{http://dx.doi.org/10.1016/0273-1177(82)90070-9}{\color{magenta}Advances in Space Research}, \href{https://ui.adsabs.harvard.edu/abs/1982AdSpR...2d.241T}{2, 241}

\bibitem[{{Vida} {et~al.}(2017){Vida}, {K{\H{o}}v{\'a}ri}, {P{\'a}l}, {Ol{\'a}h}, \& {Kriskovics}}]{Vida2017}
{Vida}, K., {K{\H{o}}v{\'a}ri}, Z., {P{\'a}l}, A., {Ol{\'a}h}, K., \& {Kriskovics}, L. 2017, \href{http://dx.doi.org/10.3847/1538-4357/aa6f05}{\color{magenta}\apj}, \href{https://ui.adsabs.harvard.edu/abs/2017ApJ...841..124V}{841, 124}

\bibitem[{{Vilhu}(1984)}]{Vilhu1984}
{Vilhu}, O. 1984, \aap, \href{https://ui.adsabs.harvard.edu/abs/1984A&A...133..117V}{133, 117}

\bibitem[{Virtanen {et~al.}(2020)Virtanen, Gommers, Oliphant, Haberland, Reddy, Cournapeau, Burovski, Peterson, Weckesser, Bright, {van der Walt}, Brett, Wilson, Millman, Mayorov, Nelson, Jones, Kern, Larson, Carey, Polat, Feng, Moore, {VanderPlas}, Laxalde, Perktold, Cimrman, Henriksen, Quintero, Harris, Archibald, Ribeiro, Pedregosa, {van Mulbregt}, \& {SciPy 1.0 Contributors}}]{Scipy2020}
Virtanen, P., Gommers, R., Oliphant, T.~E., {et~al.} 2020, \href{http://dx.doi.org/10.1038/s41592-019-0686-2}{\color{magenta}Nature Methods}, \href{https://ui.adsabs.harvard.edu/abs/2020NatMe..17..261V}{17, 261}

\bibitem[{{Wachiraphan} {et~al.}(2024){Wachiraphan}, {Berta-Thompson}, {Diamond-Lowe}, {Winters}, {Murray}, {Zhang}, {Xue}, {Morley}, {Rosario-Franco}, \& {Duvvuri}}]{Wachiraphan2024}
{Wachiraphan}, P., {Berta-Thompson}, Z.~K., {Diamond-Lowe}, H., {et~al.} 2024, \href{https://ui.adsabs.harvard.edu/abs/2024arXiv241010987W}{\href{http://dx.doi.org/10.48550/arXiv.2410.10987}{\color{magenta}arXiv e-prints}, arXiv:2410.10987}

\bibitem[{{Weiner Mansfield} {et~al.}(2024){Weiner Mansfield}, {Xue}, {Zhang}, {Mahajan}, {Ih}, {Koll}, {Bean}, {Coy}, {Eastman}, {Kempton}, \& {Kite}}]{WeinerMansfield2024}
{Weiner Mansfield}, M., {Xue}, Q., {Zhang}, M., {et~al.} 2024, \href{http://dx.doi.org/10.3847/2041-8213/ad8161}{\color{magenta}\apjl}, \href{https://ui.adsabs.harvard.edu/abs/2024ApJ...975L..22W}{975, L22}

\bibitem[{Winters {et~al.}(2021)Winters, Charbonneau, Henry, Irwin, Jao, Riedel, \& Slatten}]{Winters2021}
Winters, J.~G., Charbonneau, D., Henry, T.~J., {et~al.} 2021, \href{http://dx.doi.org/10.3847/1538-3881/abcc74}{\color{magenta}AJ}, \href{https://ui.adsabs.harvard.edu/abs/2021AJ....161...63W}{161, 63}

\bibitem[{{Wright} {et~al.}(2011){Wright}, {Drake}, {Mamajek}, \& {Henry}}]{Wright2011}
{Wright}, N.~J., {Drake}, J.~J., {Mamajek}, E.~E., \& {Henry}, G.~W. 2011, \href{http://dx.doi.org/10.1088/0004-637X/743/1/48}{\color{magenta}\apj}, \href{https://ui.adsabs.harvard.edu/abs/2011ApJ...743...48W}{743, 48}

\bibitem[{{Wright} {et~al.}(2018){Wright}, {Newton}, {Williams}, {Drake}, \& {Yadav}}]{Wright2018}
{Wright}, N.~J., {Newton}, E.~R., {Williams}, P. K.~G., {Drake}, J.~J., \& {Yadav}, R.~K. 2018, \href{http://dx.doi.org/10.1093/mnras/sty1670}{\color{magenta}\mnras}, \href{https://ui.adsabs.harvard.edu/abs/2018MNRAS.479.2351W}{479, 2351}

\bibitem[{{Xue} {et~al.}(2024){Xue}, {Bean}, {Zhang}, {Mahajan}, {Ih}, {Eastman}, {Lunine}, {Mansfield}, {Coy}, {Kempton}, {Koll}, \& {Kite}}]{Xue2024}
{Xue}, Q., {Bean}, J.~L., {Zhang}, M., {et~al.} 2024, \href{http://dx.doi.org/10.3847/2041-8213/ad72e9}{\color{magenta}\apjl}, \href{https://ui.adsabs.harvard.edu/abs/2024ApJ...973L...8X}{973, L8}

\bibitem[{{Zahnle} \& {Catling}(2017)}]{Zahnle2017}
{Zahnle}, K.~J. \& {Catling}, D.~C. 2017, \href{http://dx.doi.org/10.3847/1538-4357/aa7846}{\color{magenta}\apj}, \href{https://ui.adsabs.harvard.edu/abs/2017ApJ...843..122Z}{843, 122}

\bibitem[{{Zeng} {et~al.}(2019){Zeng}, {Jacobsen}, {Sasselov}, {Petaev}, {Vanderburg}, {Lopez-Morales}, {Perez-Mercader}, {Mattsson}, {Li}, {Heising}, {Bonomo}, {Damasso}, {Berger}, {Cao}, {Levi}, \& {Wordsworth}}]{Zeng2019}
{Zeng}, L., {Jacobsen}, S.~B., {Sasselov}, D.~D., {et~al.} 2019, \href{http://dx.doi.org/10.1073/pnas.1812905116}{\color{magenta}PNAS}, \href{https://ui.adsabs.harvard.edu/abs/2019PNAS..116.9723Z}{116, 9723}

\bibitem[{{Zhang} {et~al.}(2024){Zhang}, {Hu}, {Inglis}, {Dai}, {Bean}, {Knutson}, {Lam}, {Goffo}, \& {Gandolfi}}]{Zhang2024}
{Zhang}, M., {Hu}, R., {Inglis}, J., {et~al.} 2024, \href{http://dx.doi.org/10.3847/2041-8213/ad1a07}{\color{magenta}\apjl}, \href{https://ui.adsabs.harvard.edu/abs/2024ApJ...961L..44Z}{961, L44}

\bibitem[{{Zhao} {et~al.}(2022){Zhao}, {Fischer}, {Ford}, {Wise}, {Cretignier}, {Aigrain}, {Barragan}, {Bedell}, {Buchhave}, {Camacho}, {Cegla}, {Cisewski-Kehe}, {Collier Cameron}, {de Beurs}, {Dodson-Robinson}, {Dumusque}, {Faria}, {Gilbertson}, {Haley}, {Harrell}, {Hogg}, {Holzer}, {John}, {Klein}, {Lafarga}, {Lienhard}, {Maguire-Rajpaul}, {Mortier}, {Nicholson}, {Palumbo}, {Ramirez Delgado}, {Shallue}, {Vanderburg}, {Viana}, {Zhao}, {Zicher}, {Cabot}, {Henry}, {Roettenbacher}, {Brewer}, {Llama}, {Petersburg}, \& {Szymkowiak}}]{Zhao2022}
{Zhao}, L.~L., {Fischer}, D.~A., {Ford}, E.~B., {et~al.} 2022, \href{http://dx.doi.org/10.3847/1538-3881/ac5176}{\color{magenta}\aj}, \href{https://ui.adsabs.harvard.edu/abs/2022AJ....163..171Z}{163, 171}

\bibitem[{{Zieba} {et~al.}(2023){Zieba}, {Kreidberg}, {Ducrot}, {Gillon}, {Morley}, {Schaefer}, {Tamburo}, {Koll}, {Lyu}, {Acu{\~n}a}, {Agol}, {Iyer}, {Hu}, {Lincowski}, {Meadows}, {Selsis}, {Bolmont}, {Mandell}, \& {Suissa}}]{Zieba2023}
{Zieba}, S., {Kreidberg}, L., {Ducrot}, E., {et~al.} 2023, \href{http://dx.doi.org/10.1038/s41586-023-06232-z}{\color{magenta}\nat}, \href{https://ui.adsabs.harvard.edu/abs/2023Natur.620..746Z}{620, 746}

\end{thebibliography}
\bibliographystyle{aa_url}

\end{document}